**NICOLAS LEE GUIDOTTI**

# ACCELERATING PARTICLE-MESH ALGORITHMS WITH FPGAS AND OMPSS@OPENCL

**SÃO PAULO**

**2021**

**NICOLAS LEE GUIDOTTI**

# ACCELERATING PARTICLE-MESH ALGORITHMS WITH FPGAS AND OMPSS@OPENCL

**Undergraduate Thesis presented to Escola Politécnica da Universidade de São Paulo to obtain the degree of Electric Engineer with emphasis in Computation**

**Supervisor: Prof. Dr. Edson Midorikawa**

**SÃO PAULO**

**2021**

INDEX CARD

# ABSTRACT

Due to its flexible architecture, FPGAs support unique, deep hardware pipeline implementations for accelerating HPC applications. However, these devices are quite new in the HPC space, and thus, have been scarcely explored outside some specific scientific domain, such as machine learning or biological sequence alignment. The objective of this thesis is to characterize the FPGA-based solution for accelerating particle-mesh algorithms, in which the force applied to each particle is computed based on the fields deposited in a finite mesh (or grid). Our starting point is a 2D kinetic PIC plasma simulator called ZPIC that has the same core algorithm and functionalities as OSIRIS. To create an efficient hardware design, the program keeps the particles strictly sorted by tiles (a group of cells) and uses the local memory as an explicitly managed cache. We also create multiple copies of the local current buffer to solve dependencies during the deposition phase. The resulting pipeline was replicated multiple times to explore data parallelism and increase its throughput. We then compare our hardware solution against similar implementations on GPU and multicore CPUs, showing promising results in term of power efficiency and performance.

# RESUMO

Devido a sua arquitetura flexível, as FPGAs suportam a criação de *pipelines* personalizadas para acelerar aplicações de alto desempenho. No entanto, esses aceleradores são relativamente recentes no espaço da computação de alto desempenho (HPC), e por isso, foram pouco explorados fora de alguns domínios específicos, como *machine learning* ou alinhamento de sequências biológicas. O nosso objetivo é caracterizar uma solução baseada em FPGA para acelerar algoritmos do tipo *particle-mesh*, no qual a força aplicada em cada partícula é calculada com base no campo depositado em uma malha finita. O ponto de partida dessa monografia é um simulador cinético de plasma 2D chamado ZPIC. Para criar um projeto de hardware eficiente, o programa mantém as partículas estritamente ordenadas por *tiles* (um grupo fixo de células) e utiliza a memória local como *cache*. Nós também criamos múltiplas cópias do *buffer* da corrente local para evitar dependências durante a fase de deposição. A *pipeline* resultante foi replicada algumas vezes para explorar o paralelismo de dados e aumentar o desempenho da FPGA. Em seguida, a nossa solução em hardware foi comparada com implementações similares em GPUs e CPUs, mostrando resultados promissores em termos de eficiência energética e de desempenho.

**Palavras-chave**: FPGA, OpenCL, Simulação Cinética de Plasmas.

# LIST OF FIGURES

# LIST OF TABLES

# ACRONYMS

**ALM**      Adaptative Logic Module

**API**      Application Programming Interface

**ASIC**      Application-Specific Integrated Circuit

**BRAM**      Block RAM

**CPU**      Central Processing Unit

**CU**      Compute Unit

**DDR**      Double Data Rate

**DSP**      Digital Signal Processor

**FPGA**      Field-Programmable Gate Array

**GPU**      Graphics Processing Unit

**HBM**      High-Bandwidth Memory

**HDL**      Hardware Description Language

**HLS**      High-Level Synthesis

**HPC**      High-Performance Computing

**HT**      Intel Hyperthreading

**I/O**      Input/Output

**II**      Initiation Interval

**LUT**      Look-Up Table

**MLAB**      Memory Logic Array Block

**MPI**      Message Passing Interface

**OpenCL**      Open Computing Language

**PIC**      Particle-In-Cell

**SIMD**      Single Instruction Multiple Data

**SM**      Streaming Multiprocessor

**SWI**      Single-Work Item (Kernel)

# CONTENT

# 1    INTRODUCTION

For many years, improvements in computer hardware were driven by Dennard's power scaling (DENNARD et al., 1974) and Moore's Law (MOORE, 1998). However, with the collapse of the power scaling in 2006 and the slowdown of Moore's Law, the performance and energy efficiency of traditional processors barely changed over the last 10 years, motivating many to seek alternative designs. Among them, GPUs have been the most popular for accelerating HPC applications due to their very high throughput at a relatively low cost. However, these devices are still bound to Moore's Law and can consume up to 300W. When employed in large HPC systems, the high power consumption of these devices may expose limitations in the power delivery and cooling solutions (HENNESSY; PATTERSON, 2017; PLATT, 2018).

Recently, FPGAs emerges as a more energy-efficient option for accelerating HPC applications. Originally, FPGAs were designed for prototyping digital systems and providing low-power, embedded solutions and had fairly limited computing capabilities. These devices were also traditionally programmed using a Hardware Description Language (HDL), namely Verilog and VHDL. However, HDLs are specialized in describing logic circuits and follows a vastly different model compared to the standard programming languages, such as C/C++ and Python. For this reason, the adoption of FPGAs among software developers has been quite low.

As a way to improve FPGA usability by software programmers, manufacturers developed HLS (High-Level Synthesis) tools that convert a high-level code in C/C++ or OpenCL into a hardware description and then generate the corresponding bitstream (INTEL, 2020c; KHRONOS GROUP, 2020; XILINX, 2018). Nowadays, the programmer can even use the OpenARC compiler (LEE; VETTER, 2014a, 2014b) to transform OpenACC directives (OPENACC ORGANIZATION, 2018) into OpenCL kernels. The compiler then calls the appropriate HLS tool to generate the corresponding FPGA bitstream.

In 2014, Microsoft's Catapult project (PUTNAM et al., 2014) was the first large scale adoption of FPGAs as accelerators. With an FPGA-based solution, Microsoft significantly improved the performance of the Bing web search engine with minimal modification to their data centres due to the low power consumption, latency and space requirements of these devices. Later, both Intel/Altera and Xilinx introduced FPGAs with dedicated units for floating-point computations (INTEL, 2020a), increasing their potential for accelerating HPC applications.

Since FPGAs are quite new in the HPC space, these devices are scarcely studied outside specific scientific domains, such as neural networks (QIU et al., 2016; WANG et al., 2017), biological sequence alignment (SALAMAT; ROSING, 2020), signal processing (CHOI et al., 2003) and financial simulations (TIAN; BENKRID, 2010). Therefore, the main contribution of this thesis is *the characterization of an FPGA-based solution for accelerating particle-mesh algorithms*.

In this class of algorithms, the program computes the force applied to each particle in the system based on the fields deposited in a finite mesh (or grid) instead of calculating the particle interaction directly. In this way, the program complexity scale with $O\left(N_p\right)$ rather than $O\left(N_p{}^2\right)$ ($N_p$ is the number of particles in the system), while still producing correct results as long as the simulation is dominated by long-range interactions (i.e., longer than the mesh cell size). As particles can have any physical representation (e.g., atoms, stars or fluid components), particle-mesh algorithms are applicable to a wide range of scientific domains, ranging from fluid simulations and molecular dynamics to astrophysics. In this thesis, we focus on kinetic particle-in-cell (PIC) plasma simulation, but our findings can be extended to other particle-mesh algorithms as well.

Our starting point is the ZPIC Educational Suite (CALADO et al., 2017), a collection of simple, yet accurate PIC codes that implement the same core algorithm and functionalities as the state-of-the-art plasma simulator OSIRIS (FONSECA et al., 2002). From the ZPIC suite, we choose a 2D electromagnetic code as our base, serial implementation. This code uses a relativistic finite-difference model. For simplicity, we will refer to this program as ZPIC.

To achieve our objective, we parallelize ZPIC for three different platforms – multicore CPU, GPU, and FPGA – and compare the resulting performance, power consumption and energy efficiency. We use OpenMP (OPENMP ARB, 2018) and OpenCL (KHRONOS GROUP, 2020) to target multicore CPU and accelerators, respectively. All the host configuration and memory management are automatically handled by the OmpSs programming model (BSC PROGRAMMING MODELS, 2019; DURAN et al., 2011).

## 1.1 Thesis Outline

This document is organized as follows:

- Chapter 2 presents the general FPGA architecture and the parallelism available in these devices. Then it describes the OpenCL programming model and its application on FPGAs as well as possible optimizations. Finally, the chapter presents an overview of OmpSs and its integration with OpenCL.
- Chapter 3 provides the background theory for kinetic plasma simulations and the PIC method. This chapter also presents the ZPIC code and related works.
- Chapter 4 describes our implementation in all three platforms, highlighting the challenges encountered during the development and their solutions.
- Chapter 5 presents the performance and power efficiency results for all implementations as well as the testing methodology and the validation process.
- Chapter 6 concludes this document and presents possible future work.

# 2    PROGRAMMING WITH ACCELERATORS

## 2.1  FPGA

FPGAs, or Field-Programmable Gate Arrays, are reconfigurable devices that are able to implement custom digital circuit through the combination of basic logic blocks. Regarded as the middle of the road between general-purpose processors and ASICs (Application-Specific Integrated Circuit), the reconfigurable nature of FPGAs make them more flexible than ASICs (at the cost of higher power consumption and area usage) and more energy efficient and specialized than general-purpose processors (at the expense of more complex programming).

### 2.1.1  Architecture

Although the basic FPGA architecture is fixed, it is mostly composed of LUTs (Look-Up Tables), registers and programmable routing. Thus, these devices can implement different logics by just changing the LUT content and the routing configuration. In addition to LUTs, modern FPGAs also contain specialized blocks such as DSPs (Digital Signal Processor), memory blocks (BRAM) and different I/O controllers (for the dedicated RAM, PCIe, etc.). Besides being commonly used, these components occupy less space than a set of LUTs implementing the same logic  (ZOHOURI, 2018).

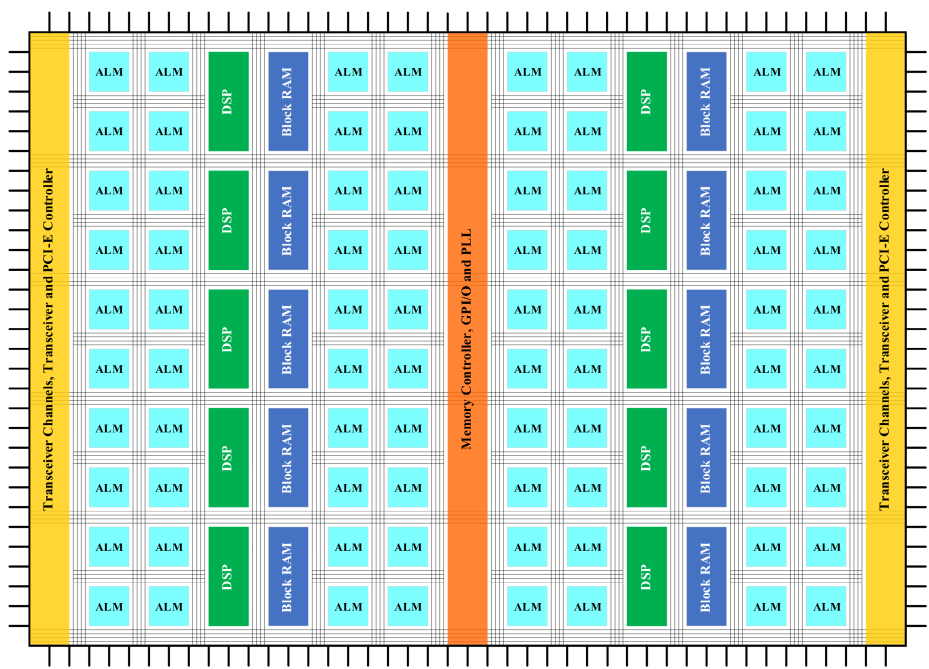

Figure 2.1 – Intel Arria 10 FPGA Architecture (ZOHOURI, 2018).

As our test system contains an Intel Arria 10 GX1150 FPGA, we will describe the architecture of this device next (Figure 2.1). Each Adaptive Logic Module (ALMs) consists of two (Adaptive) LUTs, four full adders and four registers (Flip-Flops). Depending on the input configuration, a single ALM can implement one or more logic functions. The DSPs in the FPGA provides IEEE-754-compliant single-precision floating-point addition and multiplication as well as fixed-point multiplication and addition. More complex arithmetic operations, such as convolution or dot product, can be implemented by chaining multiple DSPs.

The Arria 10 FPGA also provides two types of memory blocks: M20K and MLAB. Each M20K can store up to 20Kb of data with a maximum width of 40 bits. Each block has two ports that operate independently and can satisfy one read and one write operation at the same time. The M20K can double its operation frequency to support up to 4 logical ports (*double-pumping*, Section 2.4.3.2). In contrast, MLABs (Memory Logic Array Blocks) are better suited for storing wide and shallow array since each block can only store 640 bits. MLABs are also optimized for implementing shift registers, FIFO buffers and filter delay lines. Each MLAB is composed of up to 10 ALMs. Finally, Arria 10 also contains a controller for the external memory (DDR4 SDRAM), a PCIe interface and other I/O ports (INTEL, 2020b; ZOHOURI, 2018).

## 2.1.2 Parallelism on FPGAs

General-purpose processors have generic, fixed hardware to process a sequence of instructions (the program) from a finite set. This hardware is then replicated multiple times to achieve parallelism. In contrast, FPGA implements the entire code as a single, unique pipeline. Multiple stages of this pipeline are then executed in parallel to maximize its efficiency (*Pipeline Parallelism*). Programmers can also exploit hardware-level parallelism to increase the throughput of the device. In this case, only the hardware utilized by the algorithm is replicated.

To demonstrate the FPGA capabilities, consider a simple SAXPY example (Code Listing 2.1). Figure 2.2a shows the pipeline generated for the SAXPY code, which has four stages: 1) load the element $i$ from vectors $x$ and $y$; 2) multiply $x[i]$ by the constant $a$; 3) sum the result of the multiplication with $y[i]$; 4) store the result in the position $i$ of the vector $y$. As all operations are implemented in a dedicated hardware, multiple elements, one for each pipeline stage, can be processed at the same time (Figure 2.2b). Independent operations can also be executed in parallel (e.g., reading elements from the $x$ and $y$ vectors).

The *Initiation Interval* ($II$) is defined as the necessary distance, in clock cycles, between successive iterations of the loop in order to resolve all dependencies between them. In the SAXPY example, all iterations are completely independent, and thus, the $II = 1$ (i.e., the device can start processing a new element every clock cycle). If a dependency exists between iterations, the $II$ will be equal to the latency of the dependent operation(s). In this way, the device guarantees that dependent operation(s) is(are) completed before the next iteration begins executing the same set of operations. Any $II$ higher than 1 effectively creates bubbles in the pipeline, reducing its efficiency.

FPGAs can also exploit data parallelism through *loop unrolling* (Section 2.4.3.1) at the expense of higher resource consumption. In this case, the hardware utilized by the pipeline is replicated multiple times to execute a set of independent iterations in parallel. If the iterations access aligned and successive memory positions, these accesses can be combined into a single memory request.

Code Listing 2.1 – SAXPY code

```
void saxpy (float *y, const float a, float *x, const int size)
{
            for(int i = 0; i < size; i++)
        y[i] = a * x[i] + y[i];
}
```

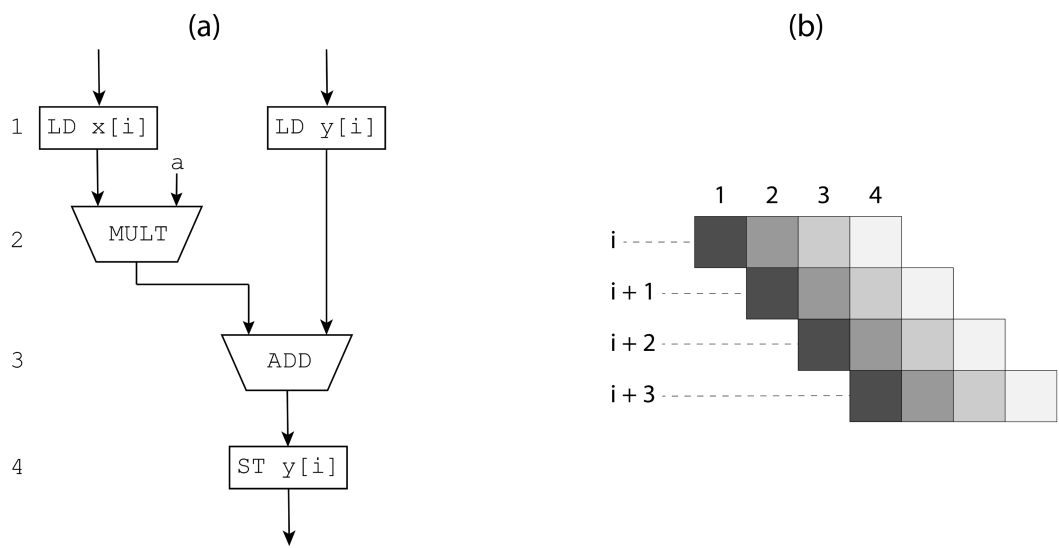

Figure 2.2 – a) Block diagram for the pipeline generated from the SAXPY code and b) its execution.

Since the entire code is implemented in hardware, not only the data can travel very fast between logic blocks but also simple logic-arithmetic operations (e.g., AND, OR, SHIFT, etc.) can be executed in less than one clock cycle. Therefore, the FPGA might be able to execute a sequence of operations within a clock cycle, while general-purpose processors take multiple cycles to process the same sequence. As an example, consider the following expression: $C = (A + B) \& 0xFF \gg 2$. In general-purpose processors, each operation in the expression is executed as a separated instruction, while the FPGA is able to execute all the operations in a single clock cycle since the adder is directly connected to a LUT that performs the AND operation. The result is then shifted 2 bits to the right by just changing the output connection.

### 2.1.3  Synthesis

Traditionally, programmers describe their FPGA designs using a Hardware Description Language (HDL), such as Verilog or VHDL. The synthesis/implementation tool (e.g., Intel Quartus (INTEL, [s.d.]) or Xilinx Vivado (XILINX, [s.d.])) then transform hardware description into a bitstream that can be loaded on the FPGA. In this process, the tool first verifies the correctness of the HDL code and then *synthesizes* it into a netlist (i.e., a list of the components of the system and their interconnections). In the next step, the functions of the netlist are *mapped* to logic blocks (e.g., DSPs, LUTs, block RAM, etc.) and *placed* on the FPGA. In the last step, the tool interconnects all the logic blocks considering some timing constraints (*routing* process). After completing all steps, the tool generates the FPGA bitstream.

Depending on the design, the complete synthesis/implementation process can take up to 12 hours or more for the Intel Arria 10 FPGA (ZOHOURI, 2018).

## 2.2 GPU

While the original purpose of the GPUs (*Graphics Processing Unit*) was to quickly render computer graphics and output the resulting image on the screen, over the years, they became powerful processors that are capable to accelerate many general-purpose applications. Modern GPUs are composed of several independent *Streaming Multiprocessors* (SMs for NVIDIA GPUs, or *Compute Units*/CUs for AMD GPUs). Each SM contains simple cores that share a register file, a RAM scratchpad, an L1 cache and a control unit. In each *wavefront* dispatched by the control unit, all 32/64 threads execute the same instruction set but operate with different data (SIMD, or Single Instruction Multiple Data). All SMs share an L2 cache and a global memory.

The device memory is divided into segments of either 32 bytes, 64 bytes or 128 bytes and aligned with their respective sizes. If a wavefront accesses one or more memory addresses within a segment, the entire segment will be requested from the device memory. Therefore, kernels achieve peak memory bandwidth when all memory accesses from the same wavefront can be *coalesced* in few memory requests as possible (e.g., 1 128-byte memory request for 32-bit words and 32 threads per wavefront). Otherwise, the data from multiple requests must be gathered in the L2/L1 cache before being delivered to the threads, increasing the instruction latency (NVIDIA, 2020).

## 2.3 OpenCL

OpenCL (KHRONOS GROUP, 2020) is an open, royalty-free standard for targeting heterogeneous architectures. Any manufacturer can support OpenCL as long as they respect the specifications and provide the necessary tools for compiling programs to their platforms. As a result, programmers can use OpenCL to target a wide range of devices, including NVIDIA/AMD GPUs, Intel/Xilinx FPGAs, Intel Xeon Phi, etc. It is worth mentioning that different architectures often require different implementations despite all accelerators supporting the same OpenCL specification.

OpenCL programs are composed of two parts: *host* and *device*. The host code handles the device configuration and data management through the OpenCL API. The host calls special functions (*kernels*) to offload computation to the device. The kernels are written in a C-like language and are executed multiple times by the device threads.

In OpenCL, the device threads are called *work-items* and are combined into *work-groups*. All work-groups and work-items are arranged into an *NDRange* grid that can have 1, 2 or 3 dimensions (Figure 2.3). The work-items of the same workgroup can be synchronized through *barriers* and can exchange information through the *local memory*. Each work-group is executed independently and there is no way to synchronize work-items from different groups.

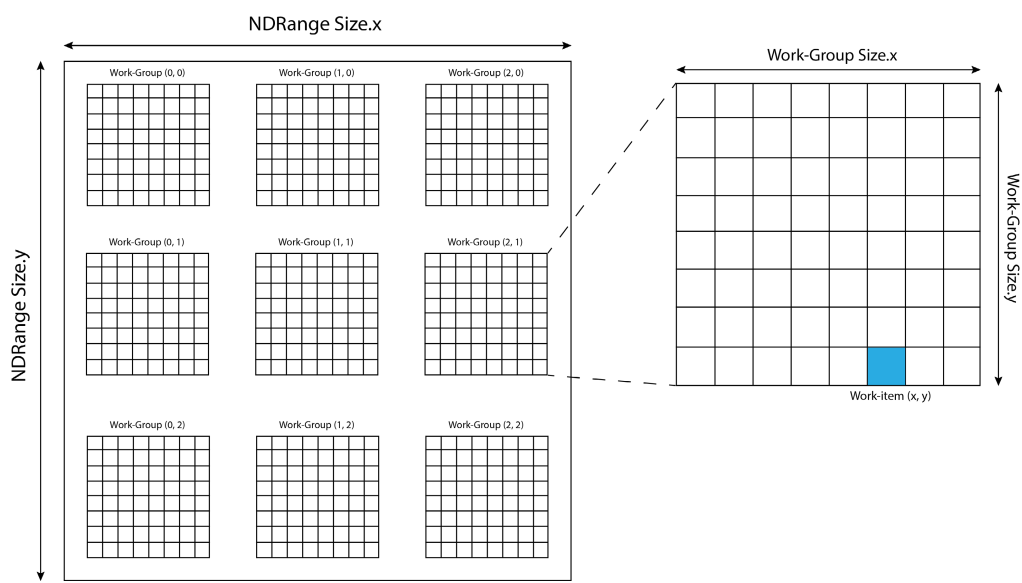

Figure 2.3 – An NDRange grid with 2 dimensions.

### 2.3.1 Memory Model

There are four memory types in the OpenCL programming model:

- **Global Memory:** The largest, but slowest memory on the device. Usually located off-chip and routed through one or more levels of cache. The global memory space is visible for all work-items in the NDRange. Global memory consistency is only guaranteed after the kernel execution is completed (KHRONOS GROUP, 2020).

- **Local Memory:** Very fast memory with limited capacity. Each work-group has an exclusive local memory space that can be accessed by all work-items of the same group. Local memory must be explicitly managed by the programmer and is usually located on-chip (e.g., RAM blocks in FPGAs). Barriers in the kernel guarantee local and global memory consistency between all work-items within the same group.

- **Constant Memory:** This memory space stores data that remains constant throughout the kernel execution. The constant memory resides in the global memory and has a dedicated cache.

- **Private Memory:** Generally resident in registers, this memory stores all data that is specific for one work-item. If the device doesn't have enough registers for executing the kernel, it spills the excess data to the global memory, resulting in a heavy performance penalty.

### 2.3.2 Execution on GPUs

The OpenCL execution model was designed primarily to exploit the architecture of the GPU and similar devices. Each work-group can be mapped directly to one SM, allowing all work-items to access the data stored in the RAM scratchpad and to synchronize their execution. In this case, the work-items are arranged in wavefronts. Although the SM can only execute one wavefront at a time, multiple wavefronts can reside in its register file. When the executing wavefront is blocked by some (memory) instruction, the SM can context-switch between the resident wavefronts and select one that is ready to be executed (i.e., is not blocked by some instruction), hiding the instruction latency of the first wavefront (VOLKOV, 2016).

The SMs execute all wavefronts in a SIMD fashion. In other words, all threads synchronously execute the same set of instructions (the kernel) over a part of the data set based on its ID in the NDRange. If the wavefront *diverges* (i.e., threads of the same wavefront follow different program flows), the SM must execute all conditional branches serially and mask the execution according to the branch that each thread follows. This divergence can severely reduce the kernel efficiency on GPUs.

## 2.4 Intel FPGA SDK for OpenCL

The Intel FPGA SDK for OpenCL provides an API, an HLS compiler and a runtime system to develop and run OpenCL programs in FPGAs. The included API not only offers extensions to the OpenCL specification (e.g., `channels`) but also finer control over FPGA design. The HLS compiler first transforms the OpenCL kernel into an HDL code and then creates the FPGA bitstream from the generated hardware description. The host code is compiled normally. During the runtime, the host loads the bitstream into the FPGA and then executes the application.

### 2.4.1 NDRange Kernels on FPGA

According to the OpenCL specification, a kernel is executed by a set of work-groups/items organized in an NDRange. In the case of GPUs, each work-group is mapped to one SM on the device. The work-items are then distributed among the cores within the associated SM (Section 2.3.2). This execution hierarchy does not exist in FPGA. Instead, the HLS compiler will create a compute unit with a deep pipeline that executes all work-items of the NDRange. Since the entire pipeline is implemented in hardware, work-items in different stages can be executed in parallel (Pipeline Parallelism, section 2.1.2). To maximize pipeline efficiency, multiple work-groups can be in-flight in the same compute unit (*work-group pipelining*). Barriers will break the pipeline in two, and each pipeline will be flushed at the barrier. There is a hardware scheduler that adjusts the initiation interval at the runtime in order to maximize the pipeline efficiency and minimize pipeline stalls.

It is possible to replicate the GPU architecture on the FPGA. With the Intel SDK, the programmer can specify the number of compute units and SIMD lanes implemented for each OpenCL kernel (Section 2.4.3.1). In this way, the work-groups will be distributed among all compute units, and each compute unit will execute multiple work-items at the same time, one on each SIMD lane (INTEL, 2020c; ZOHOURI, 2018).

## 2.4.2 Single Work-Item Kernels on FPGA

Alternatively, Intel FPGA SDK for OpenCL supports Single-Work Items (SWI) kernels, in which one work-item executes the entire kernel. All loop iterations in an SWI kernel are pipelined to achieve high performance. The compiler implements a separated pipeline for each set of nested loops. In this model, there is no hardware scheduler: the iteration scheduling is static and the $II$ is determined at the compile-time according to the dependencies between iterations of the loop. Data parallelism can be explored through *loop unrolling* (Section 2.4.3.1).

Despite all work-items of an NDRange being pipelined and no thread-level parallelism existing by default (Section 2.4.1), the OpenCL model assumes that all work-items are being executed in parallel, and thus, it is impossible to influence the scheduling distance between successive work-items in the kernel code. In this case, the thread communication is limited to the local memory, requiring barriers for synchronizing the accesses to this memory. In SWI kernels, the minimum scheduling distance between iterations is always one clock cycle. The programmer can use this information to directly transfer data from one iteration to the next as a way to solve loop-carried dependencies. There are also additional optimizations that are exclusive to the SWI kernels, such as shift registers. Based on this analysis, (ZOHOURI, 2018) concludes that Single Work-Items kernels can potentially have a lower effective initiation interval ($II$) compared to its NDRange equivalent since barriers will often be required in NDRange kernels when the local memory is utilized. Intel's Programming Guide also recommends the code to be structured as a Single Work-Item kernel (INTEL, 2020c).

Nevertheless, NDRange kernels still have some advantages over their SWI counterpart. If it is possible to achieve $II_{SWI} = 1$, SWI kernels usually have the best performance. In other situations, the compiler will determine a static scheduling distance for SWI kernels based on the longest dependency between iterations, while the hardware scheduler in the NDRange kernels will adjust the initiation interval at runtime and reorder the threads to maximize the pipeline efficiency. Therefore, the scheduler might be able to achieve a lower average initiation interval than a SWI kernel if the $II_{SWI} > 1$ (INTEL, 2020c; ZOHOURI, 2018).

## 2.4.3 Kernel Optimizations

This subsection covers optimization techniques for OpenCL-based designs on FPGAs. Some optimizations can be toggled by compilation flags, while others require significant code rewriting. All techniques presented in this section were introduced by either Zohouri or Intel's OpenCL documentation (INTEL, 2020c, 2020d; ZOHOURI, 2018).

## 2.4.3.1 Data Parallelism

There are a few ways to express data parallelism in FPGAs. The programmer can implement multiple compute units for the same NDRange kernel through the `num_compute_units()` attribute. The work-groups will then be distributed across all compute units. NDRange kernels also support the creation of SIMD lanes. Each lane replicates the pipeline stages to process multiple work-items in parallel. The memory accesses in different SIMD lanes can coalesce in a single request, increasing the memory bandwidth and reducing the area utilization. However, the number of SIMD lanes must be a power of two (to a maximum of 16 lanes) and be a divisor of the work-group size. SIMD attribute can only be used if there are no branches dependent on the work-item ID. No restriction exists for replicating the entire compute unit, but this option results in higher area utilization and lower memory bandwidth.

For SWI kernels, loop unrolling produces a similar effect to SIMD lanes in NDRange kernels. A loop can be unrolled by replicating the loop body multiple times and adjusting the control logic accordingly. If there are no loop-carried dependencies, the HLS compiler rearranges and combines the operations in the loop body in such a way that all unrolled iterations are executed in parallel.

When using SIMD lanes or loop unrolling, the compiler also coalesces consecutive and aligned memory accesses into a single wide request, improving the memory bandwidth utilization. At the same time, non-consecutive memory accesses might result in a competition for global memory bandwidth and higher replication factors in the local memory.

## 2.4.3.2 Local Memory

Differently from CPU and GPUs, there is no cache hierarchy on the FPGA. Instead, the data must be explicitly cached in the local memory to reduce the number of accesses to the global memory. Therefore, the programmer can only create efficient kernels if he/she configures and uses the local memory properly.

Registers can be used for storing temporary variables or (very) small arrays. Its access latency is only one clock cycle and a single register can be connected to multiple read/write ports. As each register can only hold one value, multiple registers must be chained together to store arrays, consuming routing resources. Depending on the array size, register chaining may be very inefficient. Large buffers are better implemented in Block RAMs (M20k and MLAB). Dynamic addressing on Blocks RAM result in access latency higher than one clock cycle and writes to this memory might create a dependency between iterations. Static addressing has a latency of one clock cycle.

Each Block RAM has only two physical ports, each one supporting independent read and write operations. The number of independent accesses can be increased to four if the Block RAM operates at double the frequency of the compute unit (*double-pumping*). Due to this limitation, the compiler often has to physically replicate the buffer to support all the concurrent accesses in a stall-free way. The FPGA resources can be easily depleted depending on the access pattern. In this case, the compiler will restart the compilation and share Block RAM ports between multiple accesses, generating an arbitration network to control the port sharing. All shared accesses are stallable.

Another option is to organize the Block RAM in banks. Each bank can operate independently from the others, offering a dedicated set of physical ports. However, this organization can only be utilized if the accesses to the bank can be determined at compile-time. Last, but not least, reducing accesses to local memory can also result in a lower replication factor. This can be achieved by using temporary registers, transposing the buffer, or even coalescing the accesses to the buffer with loop unrolling or SIMD lanes (Section 2.4.3.1).

### 2.4.3.3 Shift Registers

Block RAMs can also implement *shift registers*. In shift registers, the data is read from its head and written to its tail, with the register being shifted afterwards. In this case, the access latency is one clock cycle. Shift registers can only be inferred in Single Work-Item kernels.

Since floating-point operations in most FPGAs takes multiple cycles to be completed, a reduction operation of floating-point variables creates a data dependency between iterations, which increases the *II* of the pipeline to match the latency of the operation. According to the Intel documentation (INTEL, 2020d), this dependency can be eliminated by inferring a shift register with a size equal to the latency of the operation. With a shift register, each iteration can access a different position in the register while the floating-point operation is still in progress. In this way, the dependency is eliminated, reducing the *II* of the pipeline to one clock cycle. After the loop ends, another reduction operation over the shift register is required to obtain the final result. An unoptimized and optimized version of a dot product computation is shown in Code Listing 2.2 and 2.3, respectively.

Code Listing 2.2 – Unoptimized dot product computation.

```
float result = 0.0f;
for(int i = 0; i < size; i++)
    result += vec1[i] * vec2[i];
```

Code Listing 2.3 – Optimized dot product computation with shift registers.

```
#define FADD_LATENCY 3 // Latency of the floating-point operation

float shift_reg[FADD_LATENCY] = {0.0f};
float result = 0.0f;
for(int i = 0; i < size; i++)
{
    //Read from the head, sum and then write to the tail
    shift_reg[FADD_LATENCY - 1] = shift_reg[0] + vec1[i] * vec2[i];

    // Shift
    #pragma unroll
    for(int j = 0; j < FADD_LATENCY - 1; j++)
        shift_reg [j] = shift_reg [j + 1];
}

// Final Reduction
#pragma unroll
for(int i = 0; i < FADD_LATENCY; i++)
    result += shift_reg [i];
```

Shift registers require caution when combined with loop unrolling. Without the shift register, the floating-point operations of all unrolled iterations will be executed sequentially due to the dependency between them. Therefore, the size of the shift register must be increased to match the latency of the sequence, so each unrolled iteration can access a different position in the register, allowing them to be executed in parallel. Depending on the unroll factor, the shift register can consume a significant portion of the area resources. A better alternative is to manually unroll the loop. In this case, the original loop is split in two: the outer loop iterates over batches, and the inner loop, fully unrolled, iterates over all elements of a single batch. Since fully unrolling the loop effectively removes all loop-carried dependencies, the shift register is only required for the outer loop. And thus, it is possible to maintain the shift register size, while exploring data parallelism.

Shift registers also excel in applications that require moving a set of computation points over a regular grid, such as stencil computation, image filtering, convolutional neural network (CNN), and sequence alignment.

### 2.4.3.4 Compiler Flags

The Intel FPGA SDK for OpenCL provides several compiler flags to alter the final design and the compilation process. The most commonly used flags are shown in Table 2.1.

Table 2.1 – Compiler flags for Intel FPGA SDK for OpenCL.

| Flags | Description |
|---|---|
| `-ffp-reassoc` | The compiler implements a sequence of floating-point operations as a balanced tree, which leads to a more efficient hardware implementation at the expense of slight numeric variations. This behaviour can be controlled with `pragma fp reassoc`. |
| `-ffp-contract=fast` | Remove all intermediary floating-point rounding operations and conversions whenever possible, carrying additional bits to maintain precision. In this case, the sum and multiply are fused in a single unit (FMA, Fused Multiply-Add). This can be controlled with `pragma fp contract`. |
| `-fast-compile` | Reduce the optimization efforts to decrease compilation time. However, this feature might cause performance degradation. Only a selected number of devices support this feature. |
| `-high-effort` | Sometimes, the compiler is unable to fit the design on the device, causing the compilation to fail. With `-high-effort`, the compiler will make three attempts to recompile and generate the hardware. |
| `-incremental` | Enable an incremental compilation. The compiler will reuse the previous bitstream and only compile the modified kernels. Each subsequent compilation might degrade the performance. |
| `-profile` | The compiler inserts profile counters into the generated hardware. It will result in slightly higher resource consumption and can reduce performance. |

Source: (INTEL, 2020c).

## 2.5 OmpSs

The OmpSs programming model (BSC PROGRAMMING MODELS, 2019; DURAN et al., 2011) extends the OpenMP specification (OPENMP ARB, 2018) to support asynchronous tasks with data dependencies and heterogeneous architectures. Currently, only the first generation of OmpSs have integration with OpenCL.

Despite being an extension of OpenMP, OmpSs diverges from the standard in several key points. Instead of following OpenMP's fork-join model, OmpSs uses a thread pool to execute the tasks. In this model, the runtime first initializes the parallel environment and then assign the main task (i.e., a task with the main function) to a thread. Whenever a thread encounters a part of the code that was annotated as a `task`, the thread will instantiate it and continue the execution after the task construct. If the instantiated task can be executed immediately (e.g., all data dependencies are satisfied), the runtime will insert the task into the execution queue. The enqueue tasks will then be distributed to the available threads in the pool.

In most cases, a task requires some input data to perform some operation and produce new data that will be later used by another task or other part of the program. We define these relations as *data dependencies*. In OmpSs, there are three basic types of data dependencies: `in` (read), `out` (write) and `inout` (read/write). Based on this information, the runtime infers the relationship between tasks and creates a dependency graph. Using this graph, the runtime can determine if the task is ready to be executed or must be deferred to a later time. When a task completes, the graph is updated, allowing dependent tasks to be scheduled for execution. Notice that two tasks can be executed in parallel as long as no data dependencies exist between them. Besides the data dependencies, the tasks can be synchronized by an explicit barrier (`taskwait`).

Another point of divergence is the support for hardware accelerators. Since OpenMP version 4.5, the programmer can mark code blocks as accelerator code. The compiler will then analyse the code block and generate the kernel in the appropriate language (e.g., CUDA). In turn, in the OmpSs model, the programmer writes the kernels in the native language (CUDA, OpenCL or OpenACC) and wraps them into tasks, The device tasks are then integrated into the runtime system, following the same rules for synchronization and execution as normal tasks (SAINZ et al., 2014).

The reference implementation of OmpSs is composed of two parts: the *Mercurium* compiler and the *Nanos* runtime. Mercurium is a source-to-source compiler that identifies the OmpSs directives in the source code and substitutes them for the appropriate calls to the Nanos runtime library. Afterwards, Mercurium calls the native compiler to generate the final executable binary. Mercurium also identifies kernels written in accelerator specific languages and invoke the appropriate compiler for these files (e.g., `nvcc` for CUDA kernels). Nanos runtime provides support for the parallel execution of the application, including task instantiation and scheduling as well as data dependencies analysis.

### 2.5.1 OmpSs@OpenCL

A major weakness in the OpenCL programming model is code productivity. Since the host API is fairly low-level, the host program must include a great quantity of boilerplate code just to set up the device environment, manage the data in the accelerator and launch kernels on the device. To illustrate this issue, consider a simple SAXPY kernel (Code Listing 2.4). Even with C++ that provides a higher abstraction than standard C, there is a considerable fraction of the host code dedicated to the device and kernel setup (Code Listing 2.6).

When integrated with OpenCL, the OmpSs runtime manages and configures all accelerators transparently from the programmer. The runtime also implicitly transfers data between the host and the device according to the task's data dependencies. There is a cache system to avoid unnecessary data movement between memories. A `taskwait` will flush all data from the device back to the host unless the clause `noflush` is included (PLANAS et al., 2013; SAINZ et al., 2014). Therefore, the host code can be greatly simplified when using OmpSs@OpenCL as shown by Code Listing 2.5.

Code Listing 2.4 – A simple SAXPY kernel.

```
__kernel void saxpy(__global float *y, __global float *x,
                    const float a, const int size)
{
    const int id = get_global_id(0);
    const int stride = get_global_size(0);
    for(int i = id; i < size; i += stride)
        y[i] = a * x[i] + y[i];
}
```

Code Listing 2.5 – Host code for the SAXPY kernel in OmpSs@OpenCL.

```
#pragma omp target device(opencl) ndrange(1, size, 128) file(kernel.cl)
#pragma omp task in(x[0; size]) inout(y[0; size])
void saxpy(float *y, float *x, const float a, const int size);

int main(int argc, char **argv)
{
    const int size = atoi(argv[1]);
    std::vector<float> x(size);
    std::vector<float> y(size);
    float a;
    // Init x, y and a
    saxpy(y, x, a, size);
    #pragma omp taskwait

    return 0;
}
```

Code Listing 2.6 – Host code for the SAXPY kernel in pure OpenCL.

```cpp
int main(int argc, char **argv)
{
    const int size = atoi(argv[1]);
    std::vector<float> x(size);
    std::vector<float> y(size);
    float a;
    // Init x, y and a

    // Get platform info
    std::vector<cl::Platform> platformList;
    cl::Platform::get(&platformList);

    //Create OpenCL context
    cl_context_properties cprop[3] = { CL_CONTEXT_PLATFORM,
            (cl_context_properties) platformList[0](), 0 };
    cl::Context context(CL_DEVICE_TYPE_ACCELERATOR, cprop, NULL, NULL, &error);

    // Extract devices info
    std::vector<cl::Device> devices = context.getInfo<CL_CONTEXT_DEVICES>();

    // Create device buffers for x and y
    cl::Buffer x_acc(context, CL_MEM_READ_WRITE, size * sizeof(float),
x.data(), &error);
    cl::Buffer y_acc(context, CL_MEM_READ_WRITE, size * sizeof(float),
y.data(), &error);

    // Import and build the kernel file
    std::ifstream kernel_file("kernel.cl");
    std::string source_raw(std::istreambuf_iterator<char>(kernel_file),
(std::istreambuf_iterator<char>()));
    cl::Program::Sources source(1, std::make_pair(source_raw.c_str(),
source_raw.length() + 1));
    cl::Program program(context, source);
    program.build(devices);

    // Create the kernel and set its args
    cl::Kernel saxpy_kernel(program, "saxpy", &error);
    saxpy_kernel.setArg(0, y_acc);
    saxpy_kernel.setArg(1, x_acc);
    saxpy_kernel.setArg(2, a);
    saxpy_kernel.setArg(3, size);

    // Create a execution queue
    cl::CommandQueue queue(context, devices[0], 0, &error);

    // Copy data to the device and launch the kernel
    queue.enqueueWriteBuffer(x_acc, CL_TRUE, 0, size * sizeof(float),
x.data());
    queue.enqueueWriteBuffer(y_acc, CL_TRUE, 0, size * sizeof(float),
y.data());
    queue.enqueueNDRangeKernel(saxpy_kernel, cl::NullRange, cl::NDRange(size),
cl::NDRange(128));
    queue.enqueueReadBuffer(y_acc, CL_TRUE, 0, size * sizeof(float), y.data());
    queue.finish();

    return 0;
}
```

# 3    PLASMA SIMULATION

*Disclaimer: As an extension of the master's dissertation "Harnessing the power of modern multi-core and GPU systems with tasks with data dependencies"* (GUIDOTTI, 2021)*, this section may have significant text overlap with the aforementioned document.*

Almost all the matter in the universe is in a plasma state. Stars, gaseous nebulas and entire galaxies are only visible because there are made of plasma. On Earth, the interaction between the plasma in the ionosphere and the sun radiation creates the glow of the Aurora Borealis. Another example of plasma in the atmosphere is the flash of lightning bolts. Humans also use the plasma to create neon signs, fluorescent bulbs, TVs and other devices.

*Plasma*, in physics, is an ionized gas in which electrons are stripped from the atoms, leaving them positively charged (i.e., an ion). In most cases, gases suffer ionization when exposed to high temperatures in a similar way as heat transform solids into liquids. Therefore, many consider plasma the fourth state of the matter, alongside solids, liquids and gases. Plasmas are very susceptible to electromagnetic fields due to its charged nature (CHEN, 2016, chap. 1; GIBBON, 2016).

In technical terms, *"plasma is a quasineutral gas of charged and neutral particles which exhibits collective behaviour"* (CHEN, 2016, p. 2). The quasineutrality indicates that the total charge of the plasma is (almost) neutral. While the collective behaviour denotes that local disturbances in the equilibrium can also affect particles in remote regions due to the long-range nature of the Coulomb potential (GIBBON, 2016).

As a result, the global dynamics of the plasma is mostly determined by its internal properties (e.g., electromagnetic fields, boundary conditions, etc.), including the reaction to externally applied fields, such as lasers and particle beams (GIBBON, 2016). Therefore, the plasma behaviour can be modelled by calculating the interaction of charged particles inside the plasma. A naïve model involves calculating the Coulomb-Lorentz force between particles directly. This approach requires $O(N_p{}^2)$ operations ($N_p$ is the number of particles in the system), and thus, is only feasible with a small number of particles, limiting the complexity of the plasma events that can be simulated (VERBONCOEUR, 2005).

## 3.1  Particle-in-Cell Method

Instead of calculating the particle interactions directly, we can approximate the results through the *particle-in-cell* (PIC) method. In this method, we discretize the simulation space into a grid of cells, and then define the value of the electromagnetic field (EM) in each cell. We also consider clouds of particles rather than individual particles. The cloud's velocity and position are still defined in the continuum space. For simplicity, we will refer to these clouds as just *particles*. Finally, we advance sequentially the particles and EM fields in discrete time steps from a set of initial conditions. Note that we must choose a time step $\Delta t$ and a grid spacing $\Delta x_i, \; i = 1, 2, \dots, n$ ($n$ being the number of dimensions) that satisfy the Courant-Levy stability criterion (1) (VERBONCOEUR, 2005). The constant $c$ denote the speed of light.

$$\Delta t < \frac{1}{c}\left(\sum_i \frac{1}{(\Delta x_i)^2}\right)^{-1/2} \tag{1}$$

To advance the simulation particles in the PIC method, we calculate the Coulomb-Lorentz force applied to each particle by interpolating the EM fields' values based on the particle position. The particle motion generates an electric current that is deposited back to the grid. Using the deposit current, we then update the values of the EM fields. The same process is repeated for all time steps.

Therefore, the first step of the algorithm is to interpolate linearly values of the EM fields from the cell where the particle is located and all its adjacent cells. The position of the particle determines the contribution of each cell. It is worth mentioning that usually the EM fields are defined in a staggered grid (KANE YEE, 1966) (Figure 3.1).

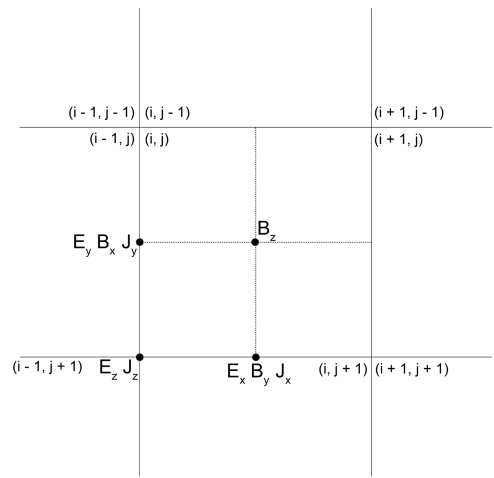

Figure 3.1 – Staggered grid for the current and electromagnetic fields (ABREU et al., 2011).

Then, we update the particle velocity and position by solving the Newton-Lorentz system of equations (VERBONCOEUR, 2005):

$$\frac{d\boldsymbol{u}}{dt} = \frac{q}{m}\left(\boldsymbol{E} + \frac{1}{c}\frac{\boldsymbol{u}}{\gamma} \times \boldsymbol{B}\right) \tag{2}$$

$$\frac{d\boldsymbol{x}}{dt} = \frac{1}{c}\frac{\boldsymbol{u}}{\gamma} \tag{3}$$

where $\boldsymbol{u}$ is the generalized velocity ($\boldsymbol{u} \equiv \boldsymbol{v}\gamma$); $\gamma$ is the Lorentz factor; q/m is the charge/mass ratio; $\boldsymbol{x}$ is the position; $\boldsymbol{E}$ and $\boldsymbol{B}$ are the EM fields interpolated at the particle's position.

$$\frac{\boldsymbol{u}_{t+\Delta t/2} - \boldsymbol{u}_{t-\Delta t/2}}{\Delta t} = \frac{q}{m}\left(\boldsymbol{E}_t + \frac{\boldsymbol{u}_{t+\Delta t/2} - \boldsymbol{u}_{t-\Delta t/2}}{2\gamma_t} \times \boldsymbol{B}_t\right) \tag{4}$$

$$\frac{\boldsymbol{x}_{t+\Delta t} - \boldsymbol{x}_t}{\Delta t} = \frac{\boldsymbol{u}_{t+\Delta t/2}}{\boldsymbol{\gamma}_{t+\Delta t/2}} \tag{5}$$

with $\gamma_t = \frac{1}{2}(\boldsymbol{\gamma}_{t+\Delta t/2} + \boldsymbol{\gamma}_{t-\Delta t/2})$.

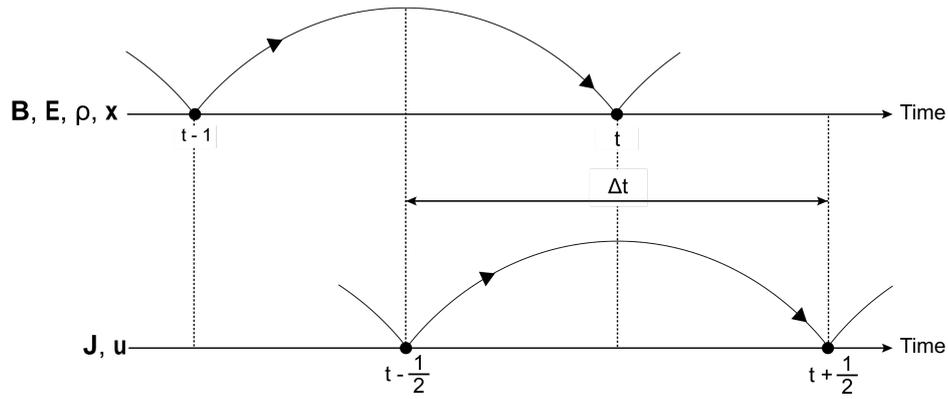

Figure 3.2 – Leapfrog scheme (BIRDSALL; LANGDON, 1991).

The Newton-Lorentz equations are often discretized using the *leapfrog* scheme (BIRDSALL; LANGDON, 1991, chaps. 15–2; VERBONCOEUR, 2005), resulting in a set of second-order difference equations (4 and 5). In this scheme, the velocity and position are updated in interleaved time steps (as shown in Figure 3.2), similar to a trajectory of a frog, hence the name. An efficient way of solving the set of difference equations is using a Boris scheme (BIRDSALL; LANGDON, 1991, chaps. 15–4; BORIS, 1970):

$$\boldsymbol{u}^- = \boldsymbol{u}_{t-\Delta/2} + \frac{q\Delta t}{2m}\boldsymbol{E}_t \tag{6}$$

$$\boldsymbol{u}' = \boldsymbol{u}^- + \boldsymbol{u}^- \times \boldsymbol{n}_t \tag{7}$$

$$\boldsymbol{u}^+ = \boldsymbol{u}^- + \boldsymbol{u}' \times \frac{2\boldsymbol{n}_t}{1 + \boldsymbol{n}_t \cdot \boldsymbol{n}_t} \tag{8}$$

$$\boldsymbol{u}_{t+\Delta t/2} = \boldsymbol{u}^+ + \frac{q\Delta t}{2m}\boldsymbol{E}_t \tag{9}$$

with

$$\boldsymbol{n}_t = \frac{q\Delta t}{2m\gamma_t}\boldsymbol{B}_t \tag{10}$$

In the Boris scheme, we first add half the electric impulse to $u_{t-\Delta/2}$ to obtain $u^-$ (6); then rotate $u^-$ with half of the magnetic impulse, resulting in $u'$ (7); next, rotate $u^-$ using the full magnetic impulse and $u'$, obtaining $u^+$ (8); and finally, add the remaining half electric impulse to $u^+$ obtaining the generalized velocity $u$ at time step $t + \Delta t/2$ (9) (BIRDSALL; LANGDON, 1991, chaps. 15–4; BORIS, 1970; VERBONCOEUR, 2005).

Using the resulting velocity, we can then update the position of the particle through equation (11).

$$\boldsymbol{x}_{t+\Delta t} = \boldsymbol{x}_t + \frac{\boldsymbol{u}_{t+\Delta t/2}}{\gamma_{t+\Delta t/2}}\Delta t \tag{11}$$

Since the motion of charged particles generates an electric current, the next step is calculating the current deposition in the grid. However, in this case, a simple interpolation of $qu/\gamma$ leads to charge conservation errors. To overcome this limitation, we need to use another approach for current deposition, such as the methods proposed by (ESIRKEPOV, 2001) and (VILLASENOR; BUNEMAN, 1992). ZPIC follows the Villasenor-Bunemann approach.

As said before, the simulation particles in the PIC method are in fact cloud of real particles, and thus, the charge is distributed on an area (in 2D or a volume in 3D). For explaining the Villasenor-Bunemann method, we will consider a uniform charge distribution over a rectangular area, as shown in Figure 3.3. Then, we use area weighting to determine the contribution from the simulation particles to the charge density of each cell.

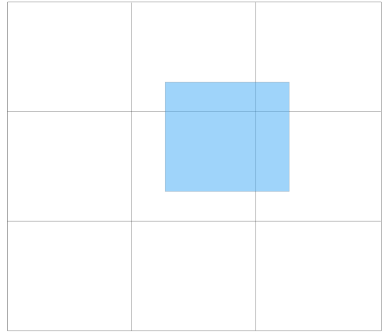

Figure 3.3 – Simple, uniform charge distribution over the cells (VILLASENOR; BUNEMAN, 1992).

In discrete cases, the current density is represented by the movement of charge between cells. As the particle moves, the intersection between the charge distribution and the cell also changes, causing charges to flow from one cell to another. From the charge conservation law (12), this charge movement is equal to the current in a given direction.

$$\nabla \cdot \boldsymbol{J} = -\frac{\partial \rho}{\partial t} \tag{12}$$

where $\boldsymbol{J}$ denote the current density; and $\rho$, the charge density.

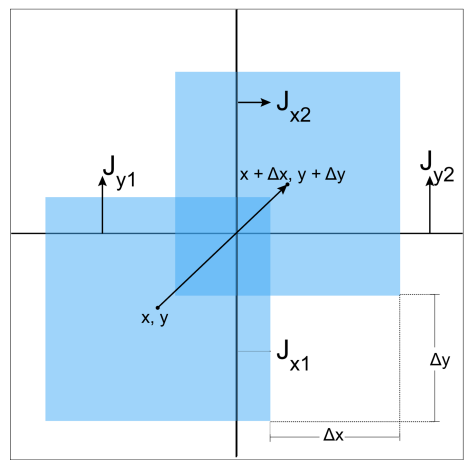

Figure 3.4 – Current generated in four cell boundaries (VILLASENOR; BUNEMAN, 1992).

Depending on the movement, a single particle can affect the current density in four, seven and even ten cell boundaries. These cases are illustrated in Figure 3.4 and 3.5. Note that all particle motion is exaggerated for illustration purposes. In the actual simulation, the time step must follow the Courant-Levy condition, mentioned previously.

The most simple and common charge movement is illustrated in Figure 3.4. As the charge area moves across the cell boundaries, it generates the currents $J_{x1}$, $J_{x2}$, $J_{y1}$ and $J_{y2}$. We define *local origin* for any charge as the nearest intersection of cell boundaries relative to the charge centre at the start of the motion. The coordinates $x$ and $y$ are relative to this local origin. Considering a unitary charge distributed in a square area with length 1, the current fluxes can be defined as follows (13-16). Notice that these equations indicate the amount of charge that crosses a cell boundary.

$$J_{x1} = \Delta x(0.5 - y - \Delta y \,/\, 2) \tag{13}$$

$$J_{x2} = \Delta x(0.5 + y + \Delta y \,/\, 2) \tag{14}$$

$$J_{y1} = \Delta y(0.5 - x - \Delta x \,/\, 2) \tag{15}$$

$$J_{y2} = \Delta y(0.5 + x + \Delta x \,/\, 2) \tag{16}$$

Sometimes, a single movement can affect seven (A) or even ten (B) boundaries (Figure **Error! Reference source not found.**). In both cases, the motion of the particle can be decomposed and analysed as multiple smaller movement that only affects four boundaries at a time. Note that the current generated in each smaller movement can affect the same boundary.

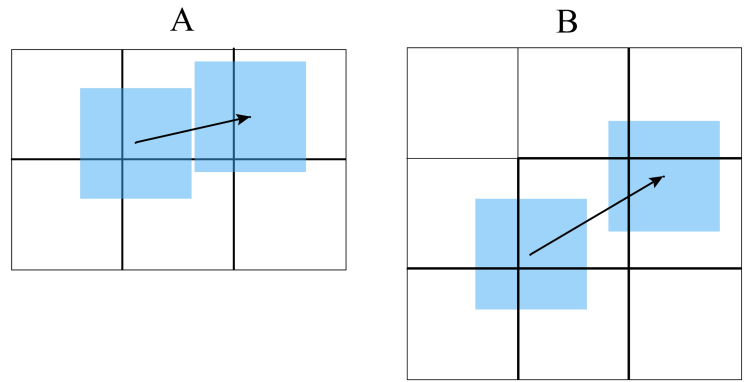

Figure 3.5 – Current generated in seven (A) and ten (B) cell boundaries (VILLASENOR; BUNEMAN, 1992).

The last step is to update the EM fields using the deposit current. By choosing appropriate units, Maxwell's equations become

$$\nabla \cdot \boldsymbol{E} = \rho \tag{17}$$

$$\nabla \cdot \boldsymbol{B} = 0 \tag{18}$$

$$\frac{\partial \boldsymbol{B}}{\partial t} = -\nabla \times \boldsymbol{E} \tag{19}$$

$$\frac{\partial \boldsymbol{E}}{\partial t} = \nabla \times \boldsymbol{B} - \boldsymbol{J} \tag{20}$$

where $\boldsymbol{E}$ and $\boldsymbol{B}$ denote the electric and magnetic fields, respectively.

First, we compute valid initial conditions that satisfy (17) and (18). Then, in each time step, we advance the electric and magnetic fields through the finite difference form of the Faraday's (19) and Ampère's (20) laws. We approximate the curl operator by using a finite-difference technique developed by (KANE YEE, 1966). As said before, the field values are defined in a staggered grid (Figure 3.1) to improve the spatial accuracy of the algorithm. Moreover, we increase the time accuracy by updating the $E$ and $B$ in three steps: first, we advance $B$ by half time step using (19); then we update $E$ by a full time step using the intermediary $B$ and Ampère's law (20); then update $B$ with the remaining half time step. Note that $E$ and $B$ are defined in integral time steps and $J$ is defined in $t + \frac{1}{2}$ time steps, following the leapfrog scheme (Figure 3.2).

## 3.2 Implementing the PIC Method: ZPIC

The ZPIC Educational Suite (CALADO et al., 2017) is a collection of bare-bones, sequential plasma simulators based on OSIRIS (FONSECA et al., 2002). Each program in the suite implements a different model/solver from the state-of-the-art plasma simulator while maintaining the same core algorithm and functionality. In this way, the ZPIC contains simple yet accurate plasma simulators that can be easily ported to other platforms and programming models. In this thesis, we focus exclusively on the 2D electromagnetic code from the ZPIC suite. The selected program follows a fully relativistic finite-difference model and the PIC method described previously (Section 3.1). For simplicity, we will refer to the program selected as ZPIC from now on.

Code Listing 3.1 Pseudocode for serial ZPIC.

```
For each time_step do
    current_zero(J)

    for each particle in simulation do
        Ep, Bp = interpolate_EMF(E, B, particle.pos)
        update_particle_momentum(Ep, Bp, particle.u)
        old_pos = particle.pos
        particle_push(particle)
        deposit_current(J, old_pos, particle.pos)
    endfor

    update_gc_add(J)

    if J_filter is enable then
        apply_filter(J, J_filter)
        update_gc_copy(J)
    endif

    yee_b(E, B, dt / 2, dx, dy)
    yee_e(E, B, J, dt, dx, dy)
    yee_b(E, B, dt / 2, dx, dy)

    update_gc_copy(E)
    update_gc_copy(B)

    if moving_window is enable then
        shift_EMF_left(E, B)
        shift_left(simulation.particles)
        insert_new_particles(simulation)
    endif
endfor
```

In the ZPIC's input file, the user not only specifies the simulation's parameters but also adds (plane and gaussian) lasers to the simulation, set a moving window (i.e., the simulation box follows the plasma events at the speed of light) or enables a binomial filter to reduce high-frequency noise. Based on this information, ZPIC will then initialize the appropriated variables and execute the simulation loop (Code Listing 3.1).

The simulation loop can be divided into two phases. In the first phase, ZPIC advances all the particles in the simulation: first, interpolating the EM field in the particle position and then updating its momentum and position. The particle motion generates an electric current that is deposited back into the grid. The second phase begins with a reduction operation (`update_gc_add`), in which the current deposited on the ghost cells (i.e., outside of the grid) are added to the corresponding cells. As an optional step, ZPIC filters the resulting current to attenuate high-frequency noise. Next, ZPIC advances the electromagnetic field through the 3-step Yee's algorithm and then copies the resulting values to the ghost cells (`update_gc_copy`). At the end of the loop, the simulation window moves one cell to the right if this feature is enabled.

To save simulation information, ZPIC introduced a new type of file format called ZDF (CALADO et al., 2017). It is lightweight (it uses data compression) and self-describing. ZPIC also includes routines to read and write in this new file format.

The ZPIC uses a unit system based on a normalized frequency $\omega_n$ similar to OSIRIS (CALADO et al., 2017; FONSECA et al., 2002):

$$t' = \frac{t}{\omega_n^{-1}} \qquad x' = \frac{\omega_n}{c}x \qquad v' = \frac{v}{c} \qquad u = \frac{\gamma v}{c} \qquad q' = \frac{q}{e} \qquad m' = \frac{m}{m_e} \tag{21}$$

$$E' = \frac{e}{m_e \omega_n c}E \qquad B' = \frac{e}{m_e \omega_n c}B \tag{22}$$

where $t$ is a time step; $x$ is the position; $v$ is the velocity; $u$ is the generalized velocity; $E$ is the electric field; $B$ is the magnetic field; $q$ is the particle charge; $m$ is the particle mass; $e$ is the absolute electron charge; $m_e$ is the electron mass; $c$ is light speed; $\gamma$ is the Lorentz factor. The apostrophe version indicates the normalized vector.

## 3.3  Related Works

SMILEI is an open-source, multi-purpose PIC code for plasma simulation that follows a patch-based overdecomposition to improve load balance and cache usage. In this approach, the simulation space is composed of many small patches (i.e., a group of cells of fixed size) organized in a Hilbert space-filling curve (HILBERT, 1935). Each MPI process contains multiple patches that are processed by the local OpenMP threads. Every couple of iterations, SMILEI verify the load distribution across the MPI processes and then exchange patches between them accordingly. Small patches sizes result in better cache efficiency as the grid quantities fit into the L1 cache (DEROUILLAT et al., 2018).

(GERMASCHEWSKI et al., 2015) developed the Plasma Simulation Code (PSC), a PIC simulator with dynamic load balancing and GPU support. PSC adopts a similar patch-based overdecomposition as SMILEI but follows a pure MPI approach. For 2D simulations, the authors adapted the PSC's algorithm to support NVIDIA GPUs with the main focus on particle-related computations. In order to use the device's shared memory as a cache, the authors implemented a modified radix sort to arrange particles in tiles (a group of cells). In this way, each thread block can load the fields into the shared memory and then push all the particles in the same tile. For the current deposition, the authors implement an atomic Villasenor-Bunemann algorithm. The authors also fused some GPU kernels in order to better utilize the available memory bandwidth.

OSIRIS is a state-of-the-art 3D plasma simulator originally developed in Fortran90 and MPI following a static domain decomposition. OSIRIS is a very mature code used by many researchers around the world and has have shown excellent scalability (up to 1.5M cores in the Sequoia supercomputer). Nowadays, OSIRIS supports an hybrid MPI + OpenMP implementation, patch-based load balancing (similar to PSC and SMILEI), GPUs (based on the algorithm proposed by (DECYK; SINGH, 2014)), Intel Xeon Phi, among other features. There is also a set of diagnostic and visualization tools (VisXD) for analysing the OSIRIS results (FONSECA et al., 2002; HEMKER, 2000).

(HARIRI et al., 2016) implemented a PIC algorithm on GPUs based on OpenACC. In the paper, the authors described different data structures for the particles (contiguous and non-contiguous layouts), presented different techniques to avoid/solve memory conflicts during the current deposition and proposed an optimized bucket sort algorithm (JOCKSCH et al., 2016) to arrange the particles in tiles. The resulting code is part of the PIC_ENGINE test bed (HARIRI et al., 2016), a PIC platform for evaluating different algorithms and methods in heterogeneous architectures.

(ABEDALMUHDI; WELLS; NISHIKAWA, 2017) studies an FPGA-based solution for improving the performance and energy efficiency of PIC plasma simulation. In their first attempt, they naively transformed the sequential code into an OpenCL kernel but it resulted in very poor performance (30x slower than the serial version). To improve their implementation, the authors used the local memory to store the grid quantities of subsets of the simulation grid, split the main loop to remove dependencies between iterations, replicated the kernel to explore data parallelism and segmented the simulation space into multiple spatial regions. The resulting code was between 2.5 and 4.0x faster than the sequential version while consuming significantly less power.

# 4    IMPLEMENTATION

Manufacturers have been unable to improve traditional processors in a significant way for the last two decades, motivating many to seek alternative designs. Among them, the GPU is the most popular, offering high raw computing power at the cost of high power consumption. In turn, FPGAs are able to create custom, dedicated hardware for the application, resulting in very energy efficient designs. In this chapter, we describe the implementation of the ZPIC in these two types of accelerators using OpenCL and OmpSs as well as a more conventional parallel approach for multicore CPU (OpenMP).

Since the FPGA resources are limited, we adopted a hybrid approach: all the particle-related routines are executed on the device, while the CPU is responsible for the grid processing (Figure 4.1). Considering that all steps of the simulation are declared as asynchronous tasks in OmpSs, device and host tasks may be executed concurrently as long as there are no dependencies between them. All FPGA kernels followed the SWI model.

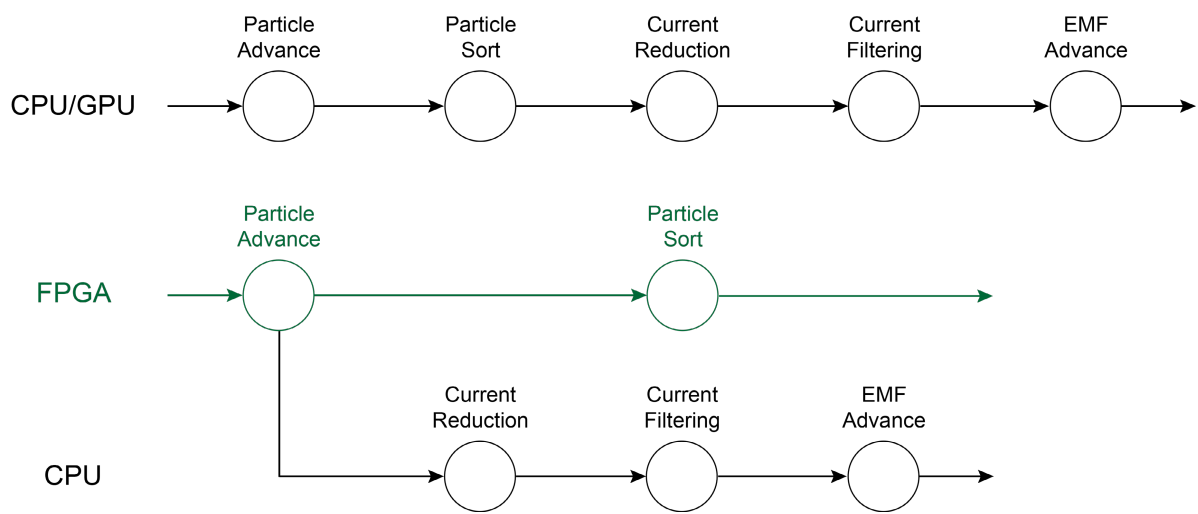

Figure 4.1 – Program flow for different platforms

## 4.1  Tiles

The main challenge when parallelizing PIC codes lies in the interactions between the particles and the grid. Particles are volatile objects moving freely in the simulation space, forcing the static grid to be accessed in random locations during the EM fields interpolation and current deposition phases. Random memory accesses break the principle of data locality that is essential for all caching techniques. On CPUs, these random accesses cause frequent cache misses and increase the latency of memory operations. On accelerators, random accesses to the global memory are particularly harmful to the program performance since they cannot be coalesced by the memory controller, decreasing the effective memory bandwidth. Without the data locality, the local memory cannot be used effectively and can even cause program crashes.

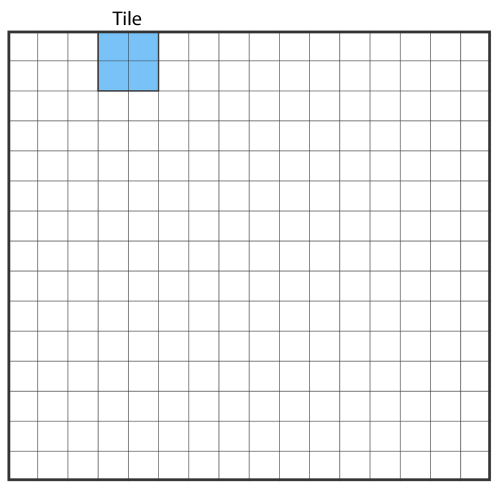

Figure 4.2 – Tile decomposition.

For this reason, many PIC codes sort the particle in regular intervals to keep spatially near particles clustered in the memory. In our case, we first decompose the grid in tiles (Figure 4.2) and then sort the particles according to the tile their located in. The grid quantities (EM fields and electric current) in each tile can be cached efficiently since the buffer size is sufficiently small to fit into either the CPU L1 cache or the device local memory. Notice that the entire grid is stored as a single continuous block in the memory and each tile reference a section of this block.

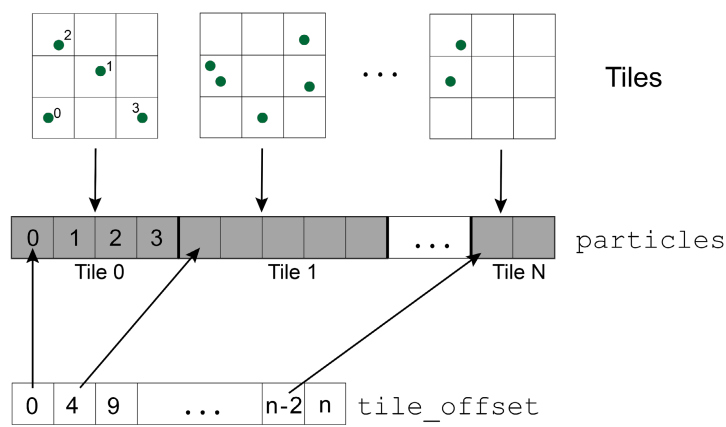

Figure 4.3 – The array section for each tile is stored in a bookmark array (`tile_offset`).

There are two ways to organize the particles in memory based on their position in the grid. The first option is to create separate arrays, one for each tile, and distribute the particles accordingly (i.e., particle *binning*). In this way, one tile is completely independent of another. However, this approach does not use the memory efficiently since the buffers must be overallocated to compensate for the concentration of particles in the space and there are gaps between them. Considering that the device memory capacity is fairly limited, we followed another approach: we use a bookmark array (`tile_offset`) to associate each tile with a section of a continuous array (`particles`) that contains all the particles in the simulation (Figure 4.3). As a downside, any changes in the number of particles in one tile affects other tiles as well (HARIRI et al., 2016; STANTCHEV; DORLAND; GUMEROV, 2008).

In realistic simulations, particles can cross the tile boundaries and access a cell outside its assigned tile. While the CPU automatically fetch the required data from the memory after a cache miss, the device local memory is treated as a user-managed cache, and thus, any out-of-bounds access to this memory often results in a crash. Therefore, the particles must be sorted every time step to preserve data locality, associating the tile and the particles within. The tile boundaries must also be extended to support particles moving to a cell outside the tile, which causes neighbour tiles to overlap.

## 4.2  Particle Advance

In a typical run, the particle advance is the most computing intensive part of the PIC algorithm as each cell contains a large number of particles (usually more than 100). This routine consists of two phases. In the first phase, the program interpolates the electromagnetic fields near the particle and then uses the interpolated fields to update the particle momenta and position. There is no direct interaction between particles in PIC simulations and each particle can be updated independently from the other, making this phase highly parallel. During the field interpolation, the program only reads data from the memory. In the second phase, the program deposits the current generated by the particle motion into the grid. However, multiple particles can generate current in the same cell, causing a data race if they are processed in parallel. Therefore, we must introduce some form of synchronization to ensure that the threads deposit the current correctly. On FPGAs, the compiler interpreted this race condition as a dependency between particles and adjusted the initiation interval accordingly.

### 4.2.1  CPU

In the CPU implementation, an OpenMP `for` loop distributes the tiles across the threads. Each thread then advances all particles within the associated tiles and deposit the current generated by the particle motion into a local, private buffer. After the particle advance is completed, the thread atomically updates the global buffer with the local values. The grid quantities of each tile are automatically loaded into the L1 cache by the CPU hardware. As the number of tiles is greater than the number of cores, we can enable a dynamic scheduling in the parallel `for` loop to improve load balance.

### 4.2.2  GPU

In the GPU implementation, each tile is mapped to one work-group in the NDRange. During the execution, the work-items of each group will load the corresponding EM fields from the global grid into the local memory, advance the particles within the tile and deposit the current generated into a local buffer. At the end of the routine, the work-items update (atomically) the global buffer with the local values. The kernel accesses the global memory in a coalesced fashion to maximize bandwidth usage (Section 2.2).

As the particles are randomly distributed within the tile, two or more work-items can advance close-by particles and deposit current in the same cell. Similarly to (DECYK; SINGH, 2014), (GERMASCHEWSKI et al., 2015) and (HARIRI et al., 2016), we found that atomic operations in the local memory the most efficient way to synchronize the work-items. However, it is still important to avoid frequent memory conflicts as the program serializes the atomic operations whenever a conflict occurs, reducing the kernel efficiency. Considering that the 32/64 particles processed by a single wavefront (i.e., one particle per thread) are less likely to update the same value at the same time if they are randomly spread inside a large tile, selecting an appropriate tile size helps to reduce the frequency of memory conflicts during the current deposition. Even though the best option is to eliminate these conflicts, reduction operations and other techniques are hard to implement on GPUs due to the limited capacity of the local memory.

Differently from CUDA, OpenCL v1.2 does not support natively atomic operation in floating-point variables, except when exchanging values in the memory. Therefore, we adapted the code proposed by (HAMURARU, 2016) to implement atomic addition for `floats` that are necessary for a correct current deposition. More recent versions of OpenCL (v2.0 and up) support C11 atomics, including floating-point addition, but very few devices support the newer standard. For example, NVIDIA GPUs are currently limited to OpenCL v1.2 with no plans to support newer revisions, while Intel FPGAs support only a subset of the features of OpenCL v2.x.

### 4.2.3 FPGA

Following the SWI model, the particle advance is implemented as a single compute unit on the FPGA (Figure 4.4). For each tile in the simulation, the compute unit loads the corresponding EM fields into the local memory (1); then advances all particles within and deposits the generated current locally (2); and finally, updates the global buffer with the local values (3). As a way to increase the compute unit throughput, stage (2) may contain multiple computing lanes operating in a SIMD-like fashion. In other words, all lanes execute the same pipeline stage for different particles in the tile. Each lane has a private copy of the current buffer to prevent data races between lanes. The lanes are created by manually unrolling the particle advance loop (Code Listing 4.1).

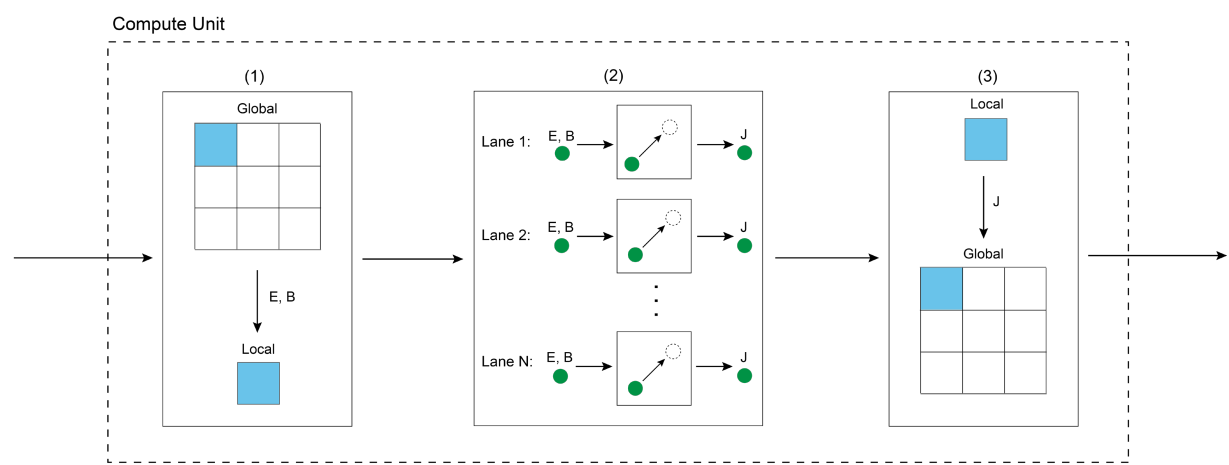

Figure 4.4 – Particle advance compute unit on the FPGA

Considering that successive particles in the pipeline can generate current in the same cell and updating a floating-point value in the local memory usually takes a few clock cycles, the minimum scheduling distance between successive particles in the pipeline (i.e., the initiation interval) is equal to the latency of the update, or 6 clock cycles – local LD (1 cycle) + FP32 FMA (4 cycles) + local ST (1 cycle). Although techniques for eliminating loop-carried dependencies exist, such as shift registers, these techniques do not apply to the PIC simulations due to the random distribution of the particles over a large tile. The unpredictability of the memory requests also prevents any type of optimization in the local memory. Instead, each access is interpreted as an unaligned, single-word request that requires an independent port, increasing the number of replicas in the local memory.

Code Listing 4.1 – Particle advance kernel with manual loop unrolling (`fpga`).

```c
#define LANES 2          // Number of pipeline replicates
#define TILE_SIZE 28     // 25 + 3 ghost cells

__kernel void spec_advance_fpga(...)
{
    for(int tile_idx = 0; tile_idx < n_tiles; tile_idx++)
    {
        int begin = tile_offset[tile_idx];
        int end = tile_offset[tile_idx + 1];

        // Local buffers
        float3 J[TILE_SIZE][TILE_SIZE][16 * LANES] = {0};
        float3 E[TILE_SIZE][TILE_SIZE];
        float3 B[TILE_SIZE][TILE_SIZE];

        load_local_EM(E, B, global_E, global_B, tile_idx);

        for(int k = begin; k < end; k += LANES)
        {
            #pragma unroll
            for(int i = 0; i < LANES; i++)
            {
                if(k + i < end)
                {
                    float3 Ep, Bp;
                    t_part part = particles[k + i];
                    interpolate_EM(part, E, B, &Ep, &Bp);
                    advance_part_momenta(part, Ep, Bp);
                    new_pos = advance_part_pos(part);
                    deposit_current(part, new_pos, J);
                    part.pos = new_pos;
                    particles[k + i] = part;
                }
            }
        }

        update_global_current(global_J, J, tile_idx);
    }
}
```

Table 4.1 - Local memory configuration for one lane in the particle advance compute unit (`fpga`).

| Local Buffer | Size | Access Latency | Banks | Replicas | Double-pumped |
|---|---|---|---|---|---|
| Magnetic Field | 28x28 | 4 | 1 | 3 | Yes |
| Electric Field | 28x28 | 4 | 1 | 3 | Yes |
| Electric Current | 28x28 | 1 | 12 | 1 | No |

Table 4.2 - Parameters for all stages in the particle advance compute unit (`fpga`).

|  | Initiation Interval | $f_{max}$ (MHz) | Pipeline Latency[1] |
|---|---|---|---|
| Load the tile's EM fields (1) | 1 | 240 | 240 |
| Main loop (2) | 6 | 240 | 513 |
| Update the global current (3) | 1 | 240 | 273 |

Table 4.1 shows the final configuration of the local memory for all grid quantities. Considering the local memory usage and program overall performance, we selected a tile size is $25 \times 25$ cells with 3 ghost cells in each dimension. Table 4.2 shows the reported parameters for all pipelines in the compute unit. The main loop corresponds to the EM interpolation, particle advance and current deposition phases of the algorithm and $f_{max}$ is the maximum operation frequency. The latency and initiation interval is measured in hardware clock cycles.

Another problem is the internal dependencies in the current deposition algorithm. The Villasenor-Buneman method splits the particle motion into 1, 2 or 3 basic movements, each one affecting four cells (Section 3.1). Differently from the general-purpose processors, the FPGA will calculate all 3 basic movements regardless of the particle motion and only select the correct result at the end. However, these movements cannot be calculated in parallel since they might affect the same cell multiple times. As a solution, we manually created 12 copies of the local current and store them in separate memory banks, so each basic movement can deposit the current in an independent copy. Afterwards, the compute unit updates the global buffer with the sum of all local copies. Notice that there are 16 copies of the current buffer per lane in Code Listing 4.1 for alignment purposes. The HLS compiler will optimize away all unused copies.

## 4.3 Particle Sorting

*Disclaimer: As an extension of the master's dissertation "Harnessing the power of modern multi-core and GPU systems with tasks with data dependencies" (GUIDOTTI, 2021), this subsection may have significant text overlap with the aforementioned document.*

On accelerators, the program uses the local memory as an explicitly managed cache for grid quantities to greatly improve memory access times. This is only possible if the program keeps the particles strictly sorted by tiles.

Fortunately, the PIC simulation is bound by the Courant-Levy condition (Section 3.1), which prevents the particle to move more than 1 cell per time step. Many authors used this restriction to create efficient sorting algorithms. (DECYK; SINGH, 2014) propose an algorithm that classifies and copies the out-of-the-order particles to a temporary array (the position in this array is determined by the particle motion). The algorithm then writes these particles back to the main array into their correct tiles, filling any holes left by the out-of-the-order particles. (GERMASCHEWSKI et al., 2015) first classifies the particle movement in 10 possible motions (4-bits) and then use a single-pass radix sort to rearrange the particle array. (STANTCHEV; DORLAND; GUMEROV, 2008) exchanges particles between nearest neighbours using the shared memory as a temporary buffer.

---

[1] The latency reported here corresponds to a kernel with a single computing lane. Replicating the computing lanes increases the pipeline latency.

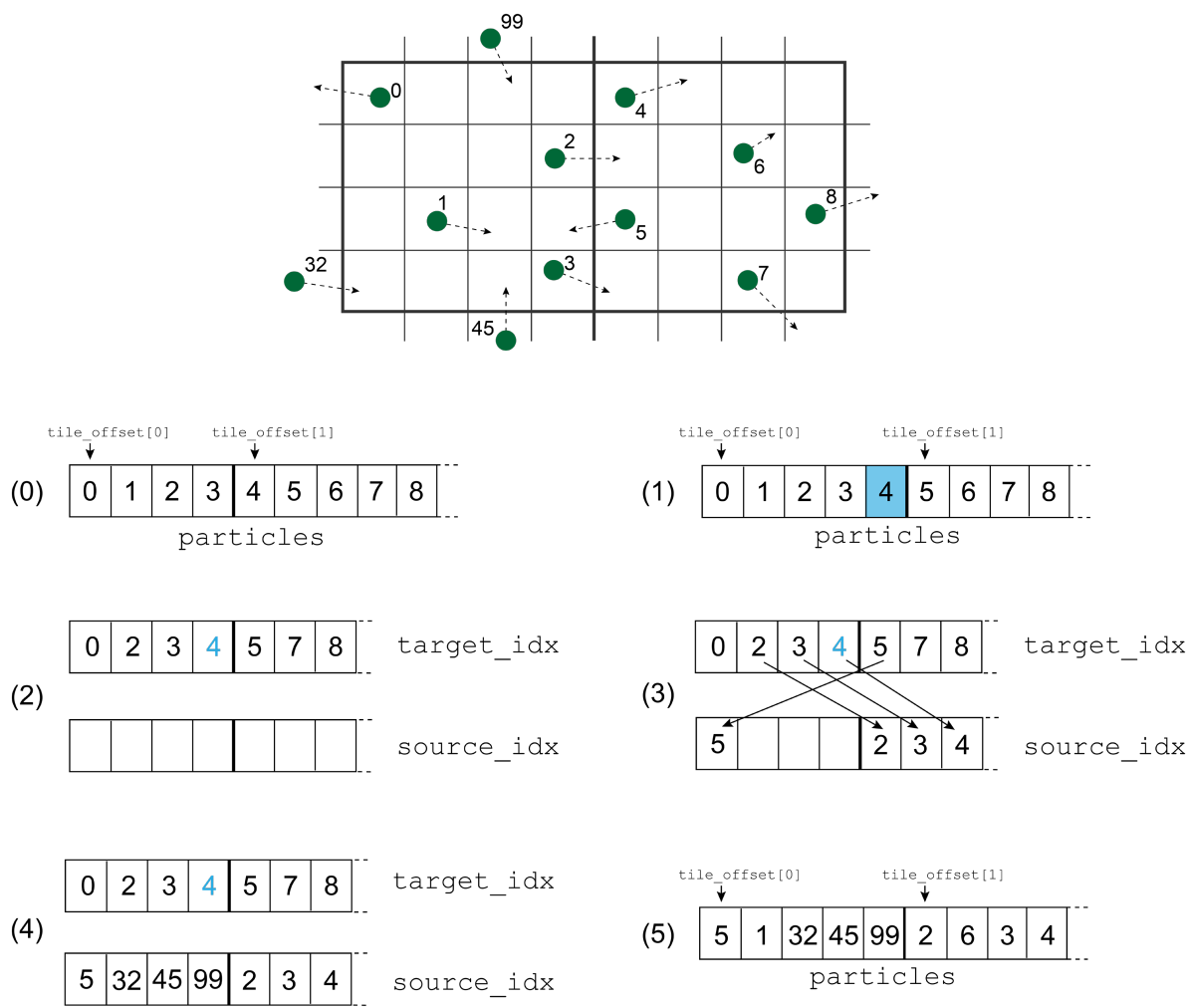

Figure 4.5 – Particle Sorting algorithm (GUIDOTTI, 2021).

In our case, we adapted the sorting algorithm proposed by (JOCKSCH et al., 2016). Depending on the tile size and the simulation conditions, only a small fraction of the particles need to the exchanged between tiles. Jocksch uses a modified bucket sort to exploit this almost sorted state, rearranging only the out-of-the-order particles (HARIRI et al., 2016). This algorithm is illustrated in Figure 4.3. The initial conditions are represented by step (0) and the particle motion, by the dashed arrows. The particles are numbered according to their index in the particle array (`particles`). A bold line separates the section in the particle array corresponding to `Tile 1` (left) and `Tile 2` (right).

During the particle sorting, there are two additional buffers: `source_idx` and `target_idx`. The `source_idx` contains the indices of all out-of-order particles, while `target_idx` stores all positions available in the particle array that these particles can move into. In other words, the algorithm will sort the array by moving all particles marked in the `source_idx` to their corresponding position indicated by the `target_idx` buffer. Each tile has a reserved section in these buffers to mark all particles entering (`source_idx`) or leaving (`target_idx`) the tile.

In the first step (1), the algorithm counts the number of particles in each tile considering the final location of each particle, then updates the bookmark array (`tile_offset`) accordingly. This bookmark array associates a section of the particle array (`particles`) with a tile in which the particles are located. Next, the algorithm registers all particles leaving their associated tile in the `target_idx` buffer (2). Notice that there are 4 particles entering `Tile 1` and 3 leaving this tile, and thus, the program must increase the corresponding section in the particle array by one position (marked in blue). Since position 4 is currently occupied by a particle in `Tile 2`, the algorithm also registers this particle in the `target_idx` buffer in order to be reallocated to a new position. In step (3), the program registers all particles entering each tile in the `source_idx` buffer, filling the "holes" in the particle array left by the exiting particles. Step (4) illustrates the final state of `source_idx` and `target_idx`. The particle array is then sorted based on these buffers (5).

Depending on the target architecture, the sorting algorithm requires some minor modifications. For example, in thread-based implementations (i.e., CPU and GPU), the particles must be registered atomically in the `source_idx` and `target_idx` buffers to prevent multiple particles to inserted at the same memory position. Table 4.3 shows the reported pipeline parameters for performing the sorting operation on the FPGA. The latency and initiation interval is measured in hardware clock cycles.

Table 4.3 - Parameters for all pipelines in the particle sorting compute unit (`fpga`).

|  | Initiation Interval | $f_{max}$ (MHz) | Pipeline Latency |
|---|---|---|---|
| Register out-of-the-order particles (2-4) | 1 | 194.4 | 524 |
| Exchange out-of-the-order particles (5) | 1 | 240 | 450 |

As the performance of both particle advance and sorting depends on the tile size, it is important to discuss the optimal size. If the tile is too small, the number of memory conflicts during the particle advance will occur more frequently (GPU), more particles will be exchanged between tiles during the particle sorting (all architectures) and each tile will contain fewer particles to amortize the latency of loading the grid quantities in the local memory (accelerators). On the other hand, very large tiles consume equally large amounts of local memory, either limiting the number of concurrent work-group that can be executed on the GPU or the number of computing lanes that can be implemented on the FPGAs. Notice that the capacity of the local memory is fairly low (e.g., AMD Radeon Instinct MI25 has 64kB of local memory per compute unit), setting a maximum tile size. On CPUs, a very large tile size can result in poor cache usage.

## 4.4  Electromagnetic Fields Advance

The parallelization of grid-based algorithms is fairly straightforward: the grid is subdivided into blocks that are distributed evenly across all threads/work-groups. Each thread/work-group then advances either the magnetic or the electric field in the assigned block. After this step is completed for all blocks, the program proceeds to the next step of the algorithm. In Yee's algorithm (KANE YEE, 1966), each cell can be processed independently from the other since to update the magnetic field in a given cell, the program only read the electric field values from neighbour cells, and vice-versa.

To improve the GPU memory latency, the kernel also follows a tile decomposition (i.e., each block corresponds to a tile in the grid). In this way, each work-group can load the tile's fields in the local memory before advancing the magnetic or electric field. To save resources on the FPGA, the EM field advance routine is executed on the CPU. Due to the OmpSs tasking model, all grid-related routines will be executed on the CPU, while the FPGA sort the particles.

# 5    Evaluation

## 5.1  Experimental Methodology

In order to verify the correctness of the parallel implementation and evaluate its performance, we simulate two theoretical, uniform plasmas as well as the collision of the two plasma clouds (Weibel Instability (WEIBEL, 1959)). Both simulations are described next.

**Uniform Plasmas:** As simple test cases, the program simulates uniform, infinite, and isolated plasmas in two specific thermal conditions: `cold` (all particles are initialized at rest) and `warm` (the particles are initialized from a thermal distribution with a width of $u_{th} = 1$). There is no initial flow velocity or EM fields. These two test cases provide ideal simulation conditions as the distribution of the particles remains uniform for the entire simulation and there is little (`warm`) to no (`cold`) particle movement. Both instances contain 25 million particles distributed over a grid of $500 \times 500$ and runs for 500 time steps.

**Collision of Plasma Clouds (Weibel):** In this simulation, the ZPIC models the collision of two plasma streams flowing in opposite directions. Both plasma streams have the same temperature and density but are composed of particles with opposite charge (electrons and positrons). The collision of the streams generates the Weibel Instability (WEIBEL, 1959) that leads to the filamentation of the plasma and the generation of the magnetic field (Figure 5.1). This simulation contains 25 million particles per species (50 million in total) distributed over a grid of $500 \times 500$ and also runs for 500 time steps.

Table 5.1 shows the hardware and software configuration to obtain all the performance results. All programs were compiled with `-O3` optimization flag. Both GPU and FPGA implementations of ZPIC uses OmpSs version 2019.06 to handle all the host side configuration and the memory transfer between the host and the device. Mercurium has GNU GCC as the back-end compiler and the Nanos++ parameters were left at their default values. The test system runs Red Hat Enterprise Linux.

Table 5.1 – Testing node specification

|  | CPU | GPU | FPGA |
|---|---|---|---|
| Model | 2x Intel Xeon Gold 5218 | AMD Radeon MI25 | Intel Arria 10 GX1150 |
| Computing Units | 2x 16C/32T | 64 CUs (4096 "cores") | - |
| Clock Frequency | 2.8 – 3.8GHz | 1.5GHz | 150 – 240MHz |
| Memory | 64GB DDR4 2677 | 16GB HBM2 | 8GB DDR4 2400 |
| Peak Memory Bandwidth | 79.47 GB/s | 464 GB/s | 34.1 GB/s |
| Software Stack | GNU GCC 8.3.1 (OpenMP v4.5) | ROCm v3.8.0 (OpenCL v1.2) | Intel Quartus v20.1 (OpenCL v1.2) |
| Kernel Compilation Flags | - | Default | `-ffp-contract=fast` `-ffp-reassoc` |

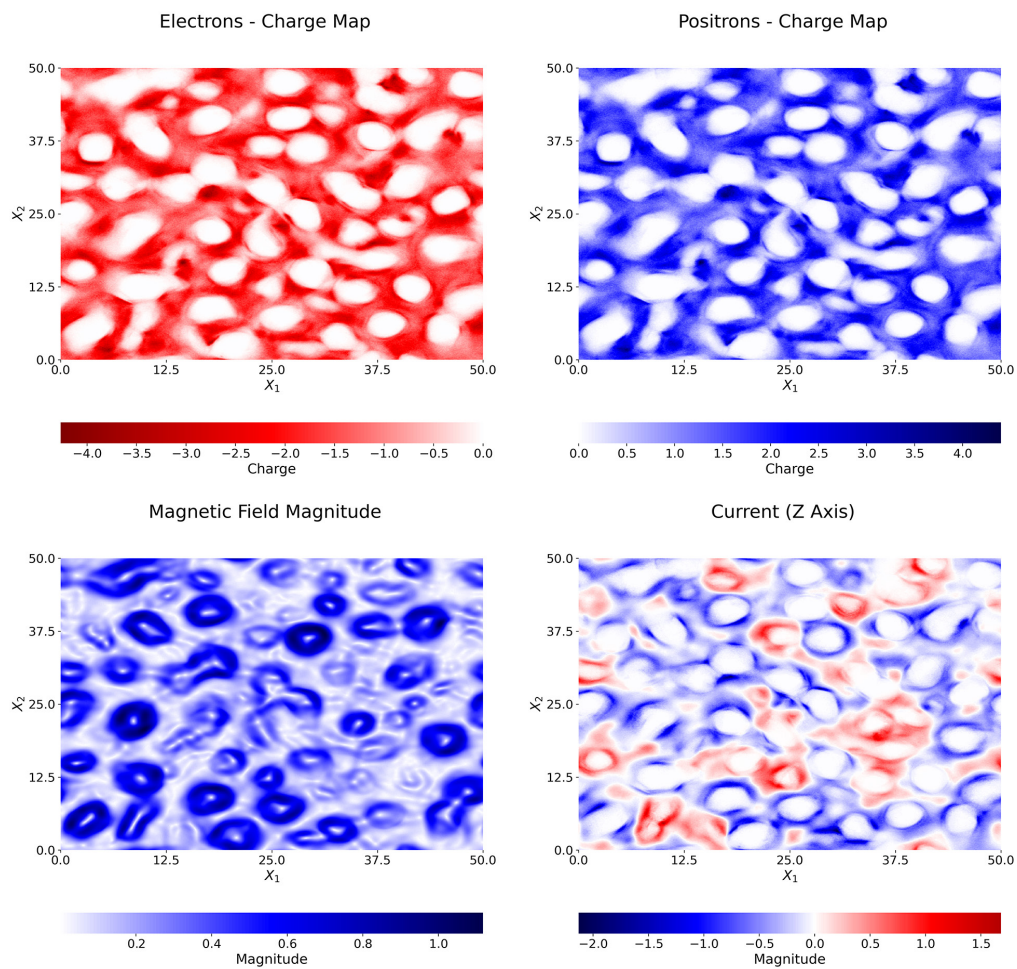

Figure 5.1 – Report for the last iteration in the `weibel` simulation

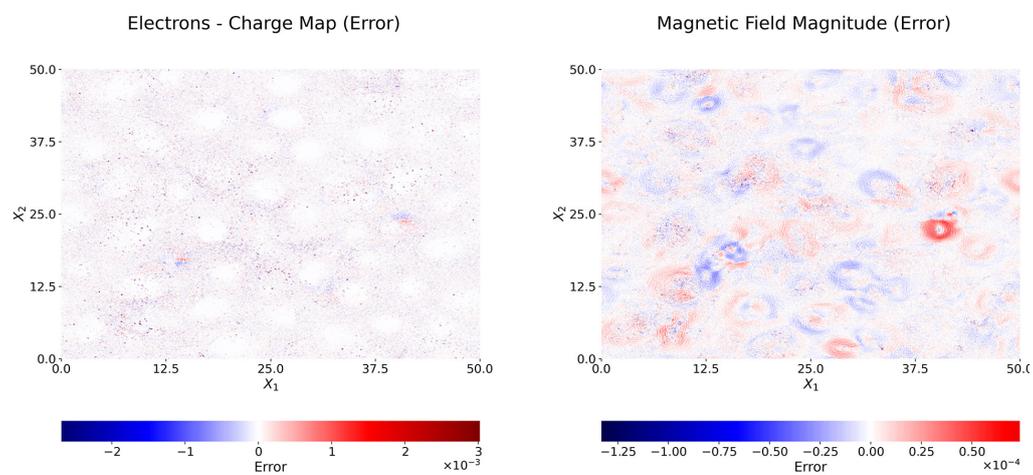

Figure 5.2 – Numerical difference between a hardware-based solution and the original implementation

## 5.2 Validation

For validating the correctness of any parallel implementation, we compare the magnetic field and charge distribution reports for the last iteration of the `weibel` simulation generated by a parallel version and the original, sequential implementation of ZPIC. The `weibel` simulation has more rigorous conditions than the uniform plasma test cases. The concurrent execution may modify the order of floating-point operations and introduce minor numerical discrepancies due to the limited precision (all implementations use FP32 calculations). Still, comparing the final reports generated by every parallel solution against the original implementation, we observer a maximum absolute error of $\sim 10^{-3}$ (Figure 5.1). This relatively small error indicates a successful implementation of ZPIC on the FPGA.

## 5.3 Performance Results

The performance results presented in this section is the average of 4 runs with a maximum relative deviation of 6.1% and 0.28% for CPU and accelerators, respectively. All speedups were calculated based on the execution time of the sequential implementation. As the CPU clock frequency varies according to the number of concurrent threads, we normalize the results considering a static frequency of 3GHz. The tiles were distributed statically in the OpenMP implementation.

### 5.3.1 FPGA

In the first set of tests, we compare our hardware-based solution against the reference implementation (Figure 5.3). Similar to the sequential ZPIC, a single "thread" executes the entire particle advance and particles sorting kernels in the FPGA implementation, pipelining all loop iterations to achieve high performance (SWI model). There are three variants of our FPGA solution, each one with a different number of computing lanes: `fpga_single` (1 lane), `fpga_dual` (2 lanes) and `fpga_quad` (4 lanes). The number of lanes only modifies the number of particles processed in each stage of the particle advance pipeline.

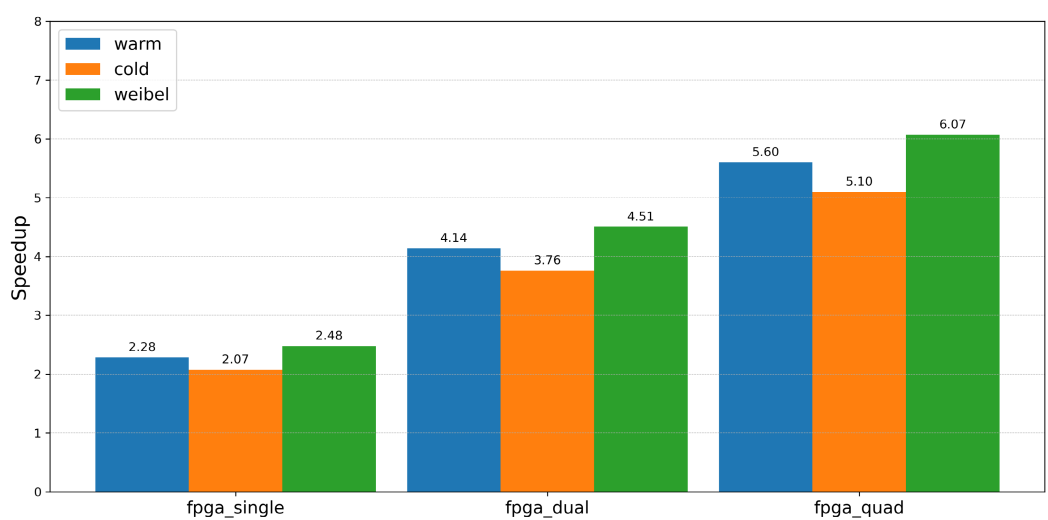

Figure 5.3 – Speedup for our solution on the Arria 10 FPGA

Although the clock frequency of the Arria 10 FPGA is 12.5x lower than the CPU and have roughly half of its memory bandwidth (and no cache hierarchy), `fpga_single` is around 2x faster than the reference implementation. On FPGAs, the entire algorithm is implemented in hardware as a unique, deep pipeline. All pipeline stages are then executed in parallel to maximize its efficiency. Each stage may have one or more operations as long as they are independent of each other. In contrast, CPUs have a fixed and short pipeline with a predefined number of ALU and load/store units. Therefore, the pipeline and instruction parallelism are fairly limited in these processors. For example, a CPU with 2 adders can only execute a maximum of 2 ADD at the same time even though the algorithm may have more than 2 independent additions. On the FPGA, all additions are implemented separately, and thus, can be executed concurrently in a single pipeline stage. Moreover, FPGA-based solutions do not have software-related overheads, such as instruction fetching and decoding, routine calls, etc.

The performance increases with the number of lanes at the cost of higher area consumption (Table 5.2). In all designs, there is a static partition on the FPGA dedicated to the interface of the device with the rest of the system and the onboard memory. Although the performance scaling of the particle advance is linear, the resulting speedup is lower since the particle sorting compute unit is the same across all variants (i.e., the sorting takes the same amount of time across all versions).

Table 5.2 - FPGA usage and speedup for different number of computing lanes

| Lanes | ALUTs | Registers | MLABs | Block RAM | DSP | Speedup |
|-------|-------|-----------|-------|-----------|-----|---------|
| 1 (single) | 16% | 17% | 3% | 21% | 19% | 1.00 |
| 2 (dual) | 18% | 18% | 4% | 40% | 37% | 1.82 |
| 4 (quad) | 23% | 22% | 5% | 56% | 73% | 2.45 |

Still, our FPGA design presents some inefficiencies that limit the overall program performance. Due to the random memory access during the particle advance, the HLS compiler is unable to optimize and combine the memory requests, resulting in higher memory access latency and replication factors. For each additional lane, the compiler must replicate the local memory subsystem, further increasing area consumption. Moreover, the initiation interval of the particle advance pipeline is quite high due to the dependencies in the current deposition. With $II = 6$ (i.e., the distance of two sets of particles in the pipeline is 6 clock cycles), the program can only maintain a maximum pipeline occupancy of $16.7\%$, drastically reducing its efficiency. As the particles generate current in a random location, it very difficult to lower this initiation interval. Notice that the $II$ is static even if successive particles rarely deposit current in the same cell.

According to the profiling results for the `fpga_dual` implementation (Table 5.3), the particle advance kernel has a major impact on the program performance, and thus, inefficiencies in this compute unit (e.g., low occupancy) lead to significant slowdowns. The profiling also demonstrates that memory-intensive routines, such as particle sorting, performs poorly on the FPGA due to the low bandwidth and lack of cache hierarchy. Last, but not least, the communication between the host and the device on hybrid applications usually impose a heavy performance loss, but it is not the case for our implementation as the tasking model permits that the FPGA sort the particles, while the CPU processes the grid quantities and performs the required data transfers. Consequently, the combined cost of communication, memory (de)allocation, device management and other overheads takes less than 3% of the execution time. The other FPGA variants show similar profiling results with different execution times for the particle advance kernel depending on the number of computing lanes.

Table 5.3 - Profiling of the `fpga_dual` implementation during the `weibel` simulation.

|  | Time (%) | Occupancy (%) | Average Stall (%) | Average Bandwidth (GB/s) |
|---|---|---|---|---|
| `particle_advance` | 71.48 | 16.3 | 0.1 | 4.151 |
| `particle_sort` | 25.78 | 90.6 | 3.1 | 6.248 |
| Others | 2.74 | - | - | - |

## 5.3.2  CPU Scaling

Next, we measure the strong scaling of the OpenMP implementation (Figure 5.4). Notice that the system has 2 CPUs with 16 physical cores each. With hyperthreading (HT) enabled, each core can issue instructions from two independent threads as a way to improve hardware efficiency. The effectiveness of hyperthreading greatly depends on the target application, and sometimes, can even decrease its performance.

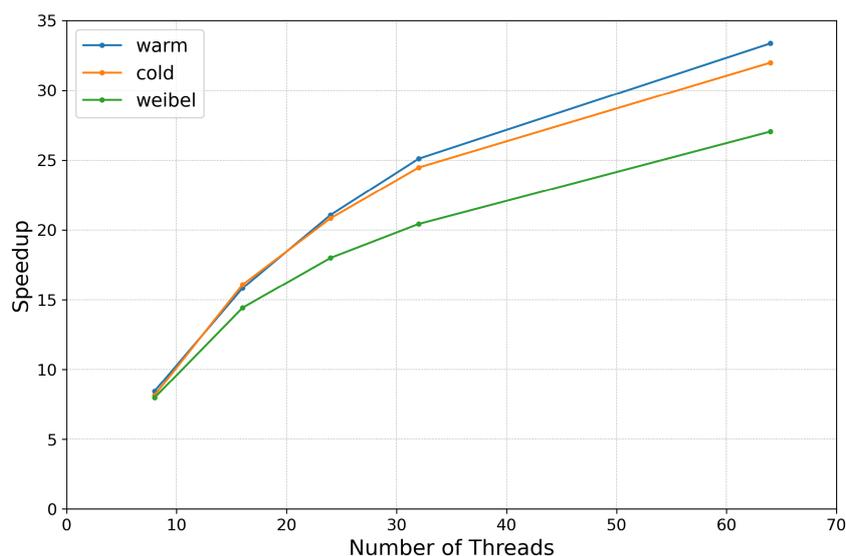

Figure 5.4 – Strong scaling for the parallel implementation (`cpu`)

Overall, the parallel program shows good scaling between 8 and 32 cores, achieving a maximum speedup of $25.10$. As expected, the parallel program has the highest speedups when simulating uniform plasmas. Both `cold` and `warm` test cases have perfect load balance across the tiles and virtually no cost for sorting the particles. In contrast, the plasma filamentation in the `weibel` simulation leads to an uneven distribution of particles over the simulation space. Due to rapid particle movement in this simulation, the program sorts a large number of particles every time step. Therefore, the parallel program has lower performance in the `weibel` test case compared to the others, but still achieves a speedup of 20.42 for 32 cores. Enabling hyperthreading improves program performance by around 30% due to the better utilization of the hardware resources.

### 5.3.3 Platform Comparison

In the subsequent test, we compare the performance of the parallel ZPIC across all architectures (Figure 5.5 and Table 5.4). Among them, the AMD Radeon MI25 GPU has the highest theoretical performance and peak memory bandwidth (HBM2 supports up to 464GB/s of memory bandwidth (AMD, 2016), while a typical DDR4 SDRAM only supports a maximum data transfer rate of 19.2GB/s per memory channel). However, only programs that are designed specifically for this architecture can harness its computing capabilities. With this in mind, we modified the ZPIC's algorithm to use the local memory as a user-managed cache, to access the global memory in a coalesced fashion and to reduce the number of memory conflicts during the current deposition as well as other optimizations. The resulting implementation is between 2.8 to 8.1x faster than its CPU counterpart with 32 cores, attain the highest speedup among all parallel platforms.

Without particle movement, the particles in the `cold` simulation do not generate electric current, and thus, there are no memory conflicts during the current deposition. As the particles remain static throughout the entire simulation, the program does not need to sort them every iteration. Consequently, the GPU performance in the `cold` simulation is vastly superior to the others. Enabling atomics as a solution for memory conflicts during the current deposition reduces the program speedup by more than half. The massive performance hit can be attributed to the lack of native support for floating-point atomics and the intrinsic cost of these operations. Even though more particles are sorted every iteration, the program achieves a higher speedup in the `weibel` simulation compared to the `warm` plasma as the higher particle entropy results in fewer memory conflicts.

Table 5.4 - Execution time (in seconds) for different implementations of ZPIC

| Version | cold | warm | weibel |
|---|---|---|---|
| serial | 1047 | 1155 | 2508 |
| cpu_16t | 72.97 | 65.15 | 173.96 |
| cpu_32t | 46.01 | 42.80 | 122.80 |
| gpu | 16.48 | 5.27 | 27.82 |
| fpga_single | 506.2 | 505.7 | 1011 |
| fpga_dual | 279.1 | 278.7 | 556.7 |
| fpga_quad | 206.2 | 205.6 | 413.2 |

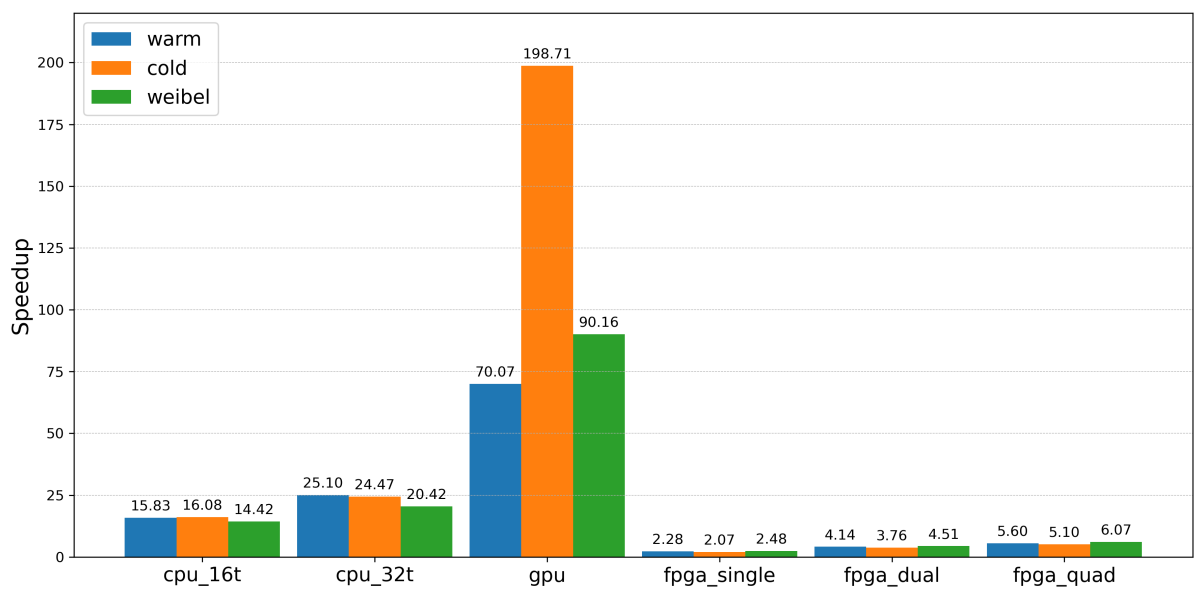

Figure 5.5 – Performance comparison between different parallel architectures

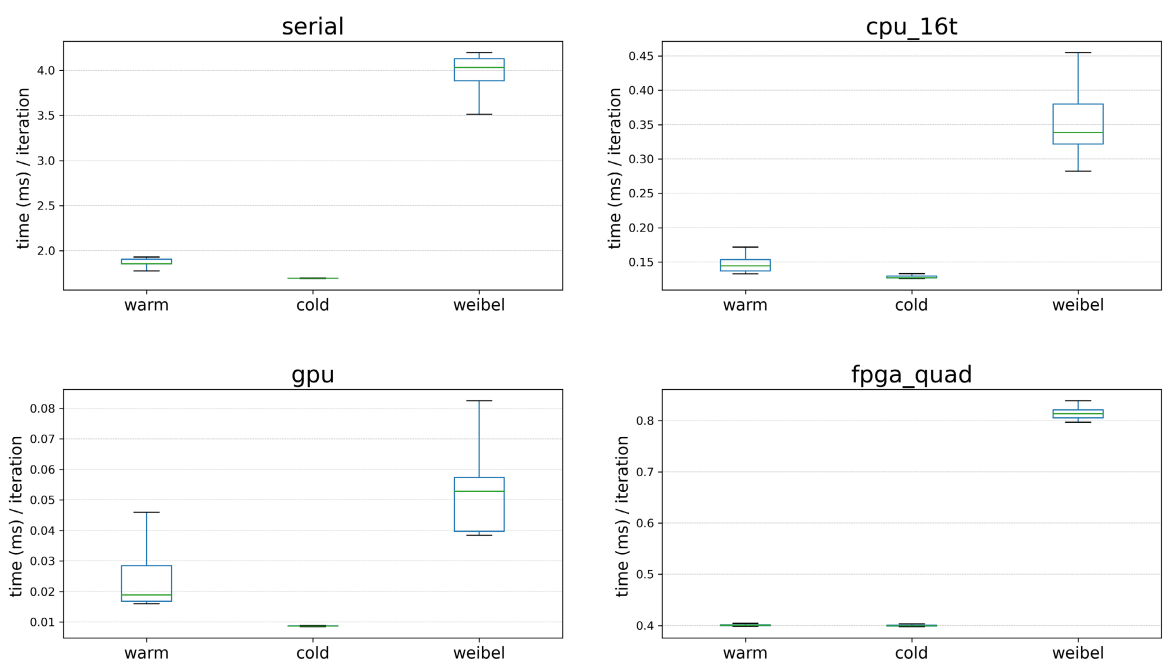

Figure 5.6 – Boxplot for the execution time per iteration for different implementations

Comparing to the other parallel platforms, the Intel Arria 10 FPGA provides a decent, but small speedup over the reference implementation. The relatively low performance can be explained by two main factors: hardware/software limitations and design inefficiencies. Hardware-wise, the Intel Arria 10 has lower clock frequency and lower memory bandwidth than the other devices, has no cache hierarchy and uses older technologies (e.g., this FPGA was launched in 2013, while the AMD MI25 was released in 2017). HLS tools also are quite recent, and thus, are not fully mature or optimized yet. Since the FPGA resources are finite, the number of lanes and compute units that can be implemented on the device are also limited. With just 4 lanes, our design already uses 73% of all DSP available on the device. In comparison, each CPU in the system has 16 cores (for a total of 32 cores), while the AMD Radeon MI25 has 4096 "cores". Finally, as stated before, our FPGA design has some limitations, such as high initiation interval, low pipeline occupancy, high replication factors and unoptimized memory accesses.

Another interesting comparison between implementations is the variation of execution time per iteration (Figure 5.6). As the particles remain static in the `cold` simulation, all time steps take the same amount of time independently of the target architecture. In the `warm` simulation, the small particle movement can slightly alter the time required for processing one iteration. The Weibel instability drastically changes the plasma behaviour, and thus, the execution time for one iteration may be significantly different from another, especially when comparing iteration from the beginning and the end of the simulation.

In the GPU, the time taken to complete one iteration heavily depends on the number of memory conflicts during the current deposition, showing the highest time variation among all versions since there are no (`fpga` and `serial`) or very few (`cpu_16t`) conflicts in the other implementations. Additionally, some iterations can be processed faster than others as modern processors dynamically adjust the clock frequency depending on power consumption, temperature, and processor load. While on FPGAs, the entire algorithm is implemented in hardware with a static clock frequency. Therefore, the performance in this device is very consistent throughout the entire simulation even for the `weibel` test case.

## 5.4  Power Efficiency

Modern CPUs and accelerators often have built-in power sensors that accessible through the manufacturer's API, driver and/or utility, such as `fpgainfo` (INTEL, 2017), `rocm-smi` (AMD, 2021) and `sysfs powercap` (IMES et al., 2019). Using these tools, we measure the power consumption and efficiency on each platform during the `weibel` simulation. For GPU and FPGAs, the program queries the power sensor every 10ms during the simulation and then average the reported values. The energy to solution is calculated by multiplying the average power consumption with the simulation run time. As CPU sensors report energy instead of power consumption, the energy to solution can be directly measured by subtracting the energy accumulated at the beginning of the execution from the one at the end. It is worth mentioning that the CPU sensor only reports the energy usage by the CPU package, while the accelerators report the consumption for the entire board. Figure 5.7 show the average power consumption and efficiency of each device.

Considering that the system contains two identical CPUs working in unison, the combined power usage naturally surpasses the other platforms, but only provides a modest speedup in return. Therefore, these processors consume a hefty amount of energy for simulating the Weibel instability. Despite almost having the same power envelope, the GPU is 6.8x more energy efficient than the CPUs due to the shorter simulation time. Since FPGAs only implements the necessary hardware and have low clock frequencies, the Intel Arria 10 consumes 6.92x less power than the CPU and are 2.68x more energy efficient. Compared to the GPU, the FPGA have 5.88x lower power consumption but loses in terms of energy efficiency due to its low performance.

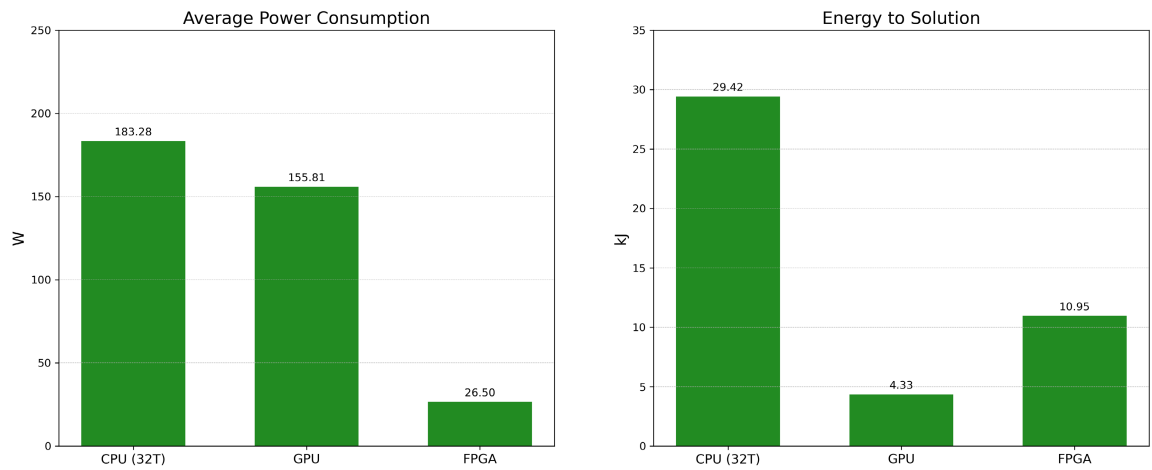

Figure 5.7 – Power efficiency of each architecture.

# 6    CONCLUSION

In this thesis, we successfully developed an FPGA-based solution for a particle-mesh algorithm that is capable of executing complex simulations with minor numerical differences from the original, sequential CPU code. In term of performance, the hardware-based solution provides decent speedup over the reference implementation, while showing better energy efficiency and performance consistency than a traditional processor. However, the FPGA performance was significantly worse than a modern multicore CPU or a GPU due to the intrinsic limitations of the device. For example, the global memory bandwidth of the FPGA is considerably lower than the other platforms and its memory controller is not as efficient (e.g., in both CPU and the GPU controllers can realign memory request to increase memory bandwidth usage, while the FPGA controller does not support this feature).

Besides the hardware and software limitations of the FPGAs, our design also has some inefficiencies. First, the random distribution of particles over the tile prevents the creation of an efficient local memory configuration. In this case, the HLS compiler interprets each random memory request as single-word, unaligned access that requires an independent port, which increases the pipeline latency and resource consumption. Additionally, as successive particles can update the same cell, the compiler must increase the initiation interval of the pipeline to match the latency of the operation. Even if these conflicts rarely occur, the initiation interval is statically determined by the worst-case scenario. A high initiation interval drastically reduces the pipeline efficiency. Based on these findings, we can conclude that FPGA is not well suited for accelerating dynamic algorithms, such as particle-mesh codes, since the unpredictability prevents the compiler to generate an optimized kernel. Still, with more research, development time and creativity, it may be possible to create an efficient hardware design due to the flexible architecture of the FPGAs.

Nevertheless, the FPGA hardware has been rapidly evolving in recent years with newer models support significantly higher clock frequencies and larger designs. Some models even incorporate newer technologies, such as HBM. As the performance grows and the HLS tools become more mature, FPGAs may become a compelling option for accelerating HPC applications.

## 6.1  Future Work

For future work, we propose some improvements to our hardware design. Dynamic scheduling, in which the pipeline initiation interval is adjusted during the runtime, can significantly improve the particle advance pipeline occupancy as memory conflicts during the current deposition are rare when considering a large tile. Alternatively, the pipeline efficiency can also be greatly improved by removing the loop-carried dependencies through either temporary variables or full copies of the grid quantities. Another possible optimization is to sort multiple particles in parallel, which may reduce the cost of this operation. Last, but not least, we propose the implementation of the entire PIC code on the FPGA as a way to explore a pure hardware solution, which enables the use of FPGA specific features, such as `channels`.

# BIBLIOGRAPHY

ABEDALMUHDI, A.; WELLS, B. E.; NISHIKAWA, K.-I. **Efficient Particle-Grid Space Interpolation of an FPGA-Accelerated Particle-in-Cell Plasma Simulation**. 2017 IEEE 25th Annual International Symposium on Field-Programmable Custom Computing Machines (FCCM). **Proceedings:**... In: 2017 IEEE 25TH ANNUAL INTERNATIONAL SYMPOSIUM ON FIELD-PROGRAMMABLE CUSTOM COMPUTING MACHINES (FCCM). Napa, CA, USA: IEEE, Apr. 2017

ABREU, P. et al. PIC Codes in New Processors: A Full Relativistic PIC Code in CUDA-Enabled Hardware With Direct Visualization. **IEEE Transactions on Plasma Science**, v. 39, n. 2, p. 675–685, Feb. 2011.

AMD. **Radeon Instinct MI25 Datasheet**. URL: <https://www.amd.com/en/system/files?file=documents/radeon-instinct-mi25-datasheet.pdf>. Accessed in: 6 mar. 2021.

AMD. **AMD ROCm v4.0.1 Documentation**. URL: <https://rocmdocs.amd.com/en/latest/index.html>. Accessed in: 10 mar. 2021.

BIRDSALL, C. K.; LANGDON, A. B. **Plasma Physics via Computer Simulation**. 1st. ed. Boca Raton: CRC Press, 1991.

BORIS, J. P. **Relativistic plasma simulation - optimization of a hybrid code**. . In: PROCEEDING OF FOURTH CONFERENCE ON NUMERICAL SIMULATIONS OF PLASMAS. 1970

BSC PROGRAMMING MODELS. **OmpSs Specification**. URL: <https://pm.bsc.es/ftp/ompss/doc/spec/OmpSsSpecification.pdf>. Accessed in: 15 feb. 2021.

CALADO, R. et al. The ZPIC educational code suite. **APS Division of Plasma Physics Meeting Abstracts**, v. 2017, p. JP11.004, Oct. 2017.

CHEN, F. F. **Introduction to Plasma Physics and Controlled Fusion**. Cham: Springer International Publishing, 2016.

CHOI, S. et al. **Energy-efficient signal processing using FPGAs**. Proceedings of the 2003 ACM/SIGDA eleventh international symposium on Field programmable gate arrays. **Proceedings:**...: FPGA '03.New York, NY, USA: Association for Computing Machinery, 23 Feb. 2003

DECYK, V. K.; SINGH, T. V. Particle-in-Cell algorithms for emerging computer architectures. **Computer Physics Communications**, v. 185, n. 3, p. 708–719, 1 Mar. 2014.

DENNARD, R. H. et al. Design of ion-implanted MOSFET's with very small physical dimensions. **IEEE Journal of Solid-State Circuits**, v. 9, n. 5, p. 256–268, Oct. 1974.

DEROUILLAT, J. et al. SMILEI: a collaborative, open-source, multi-purpose particle-in-cell code for plasma simulation. **Computer Physics Communications**, v. 222, p. 351–373, Jan. 2018.

DURAN, A. et al. OmpSs: A proposal for programming heterogeneous multi-core architectures. **Parallel Processing Letters**, v. 21, n. 02, p. 173–193, 1 Jun. 2011.

ESIRKEPOV, T. ZH. Exact charge conservation scheme for Particle-in-Cell simulation with an arbitrary form-factor. **Computer Physics Communications**, v. 135, n. 2, p. 144–153, 1 Apr. 2001.

FONSECA, R. A. et al. **OSIRIS: A Three-Dimensional, Fully Relativistic Particle in Cell Code for Modeling Plasma Based Accelerators**. (P. M. A. Sloot et al., Eds.) Computational Science — ICCS 2002. **Proceedings:**...Berlin, Heidelberg: Springer, 2002

GERMASCHEWSKI, K. et al. The Plasma Simulation Code: A modern particle-in-cell code with load-balancing and GPU support. **arXiv:1310.7866 [physics]**, 12 Nov. 2015.

GIBBON, P. Introduction to Plasma Physics. **CERN Yellow Reports**, p. 51 Pages, 16 Feb. 2016.

GUIDOTTI, N. **Harnessing the power of modern multi-core and GPU systems with tasks with data dependencies**. Lisboa: Instituto Superior Técnico, 2021.

HAMURARU, A. **Atomic operations for floats in OpenCL - improvedStreamHPC**, 9 Feb. 2016. URL: <https://streamhpc.com/blog/2016-02-09/atomic-operations-for-floats-in-opencl-improved/>. Accessed in: 7 mar. 2021

HARIRI, F. et al. A portable platform for accelerated PIC codes and its application to GPUs using OpenACC. **Computer Physics Communications**, v. 207, p. 69–82, 1 Oct. 2016.

HEMKER, R. G. **Particle-In-Cell Modeling of Plasma-Based Accelerators in Two and Three Dimensions**. Los Angeles: University of California, 2000.

HENNESSY, J. L.; PATTERSON, D. A. **Computer Architecture: A Quantitative Approach**. 6th edition ed. Cambridge, MA: Morgan Kaufmann, 2017.

HILBERT, D. Über die stetige Abbildung einer Linie auf ein Flächenstück. In: HILBERT, D. (Ed.). . **Dritter Band: Analysis · Grundlagen der Mathematik · Physik Verschiedenes: Nebst Einer Lebensgeschichte**. Berlin, Heidelberg: Springer, 1935. p. 1–2.

IMES, C. et al. **CoPPer: Soft Real-Time Application Performance Using Hardware Power Capping**. 2019 IEEE International Conference on Autonomic Computing (ICAC). **Proceedings:**... In: 2019 IEEE INTERNATIONAL CONFERENCE ON AUTONOMIC COMPUTING (ICAC). Jun. 2019

INTEL. **Open Programmable Acceleration Engine (OPAE) v2.0 Documentation**. URL: <https://opae.github.io/2.0.0/index.html>. Accessed in: 10 mar. 2021.

INTEL. **Intel® Arria® 10 Device Overview**. URL: <https://www.intel.com/content/dam/www/programmable/us/en/pdfs/literature/hb/arria-10/a10_overview.pdf>. Accessed in: 30 jan. 2021a.

INTEL. **Intel® Arria® 10 Handbook**. URL: <https://www.intel.com/content/dam/www/programmable/us/en/pdfs/literature/hb/arria-10/a10_handbook.pdf>. Accessed in: 30 jan. 2021b.

INTEL. **Intel FPGA SDK for OpenCL Pro Edition: Programming Guide**. URL: <https://www.intel.com/content/www/us/en/programmable/documentation/mwh1391807965224.html>. Accessed in: 30 jan. 2021c.

INTEL. **Intel FPGA SDK for OpenCL Pro Edition: Best Practices Guide**. URL: <https://www.intel.com/content/dam/www/programmable/us/en/pdfs/literature/hb/opencl-sdk/aocl-best-practices-guide.pdf>. Accessed in: 30 jan. 2021d.

INTEL. **Intel® Quartus® Prime Software Suite**. URL: <https://www.intel.com/content/www/us/en/software/programmable/quartus-prime/overview.html>. Accessed in: 30 jan. 2021.

JOCKSCH, A. et al. **A bucket sort algorithm for the particle-in-cell method on manycore architectures**. Parallel Processing and Applied Mathematics. **Proceedings:**... In: 11TH INTERNATIONAL CONFERENCE, PPAM 2015. Springer, 2016

KANE YEE. Numerical solution of initial boundary value problems involving maxwell's equations in isotropic media. **IEEE Transactions on Antennas and Propagation**, v. 14, n. 3, p. 302–307, May 1966.

KHRONOS GROUP. **The OpenCL$^{TM}$ Specification, Version 3.0**. URL: <https://www.khronos.org/registry/OpenCL/specs/3.0-unified/pdf/OpenCL_API.pdf>. Accessed in: 31 jan. 2021.

LEE, S.; VETTER, J. S. **OpenARC: open accelerator research compiler for directive-based, efficient heterogeneous computing**. Proceedings of the 23rd international symposium on High-performance parallel and distributed computing. **Proceedings:**...: HPDC '14.New York, NY, USA: Association for Computing Machinery, 23 Jun. 2014a

LEE, S.; VETTER, J. S. **OpenARC: Extensible OpenACC Compiler Framework for Directive-Based Accelerator Programming Study**. 2014 First Workshop on Accelerator Programming using Directives. **Proceedings:**... In: 2014 FIRST WORKSHOP ON ACCELERATOR PROGRAMMING USING DIRECTIVES. Nov. 2014b

MOORE, G. E. Cramming More Components Onto Integrated Circuits. **Proceedings of the IEEE**, v. 86, n. 1, p. 82–85, Jan. 1998.

NVIDIA. **CUDA C++ Programming Guide, Version 11.1**. URL: <https://docs.nvidia.com/pdf/CUDA_C_Programming_Guide.pdf>. Accessed in: 9 feb. 2021.

OPENACC ORGANIZATION. **OpenACC Application Programming Interface, Version 2.7**. URL: <https://www.openacc.org/sites/default/files/inline-files/OpenACC.2.7.pdf>. Accessed in: 31 jan. 2021.

OPENMP ARB. **OpenMP Application Programming Interface, Version 5.0**. URL: <https://www.openmp.org/wp-content/uploads/OpenMP-API-Specification-5.0.pdf>. Accessed in: 14 feb. 2021.

PLANAS, J. et al. **Self-Adaptive OmpSs Tasks in Heterogeneous Environments**. 2013 IEEE 27th International Symposium on Parallel and Distributed Processing. **Proceedings:**... In: 2013 IEEE INTERNATIONAL SYMPOSIUM ON PARALLEL & DISTRIBUTED PROCESSING (IPDPS). Cambridge, MA, USA: IEEE, May 2013

PLATT, S. **Metamorphosis of an industry, part two: Moore's Law and Dennard Scaling**. URL: <https://www.micron.com/about/blog/2018/october/metamorphosis-of-an-industry-part-two-moores-law>. Accessed in: 14 dec. 2020.

PUTNAM, A. et al. A Reconfigurable Fabric for Accelerating Large-Scale Datacenter Services. **IEEE Micro**, v. 35, n. 3, p. 10–22, May 2014.

QIU, J. et al. **Going Deeper with Embedded FPGA Platform for Convolutional Neural Network**. Proceedings of the 2016 ACM/SIGDA International Symposium on Field-Programmable Gate Arrays. **Proceedings:**...: FPGA '16.New York, NY, USA: Association for Computing Machinery, 21 Feb. 2016

SAINZ, F. et al. **Leveraging OmpSs to Exploit Hardware Accelerators**. 2014 IEEE 26th International Symposium on Computer Architecture and High Performance Computing. **Proceedings:**... In: 2014 26TH INTERNATIONAL SYMPOSIUM ON COMPUTER ARCHITECTURE AND HIGH PERFORMANCE COMPUTING (SBAC-PAD). Jussieu, Paris, France: IEEE, Oct. 2014

SALAMAT, S.; ROSING, T. FPGA Acceleration of Sequence Alignment: A Survey. **arXiv:2002.02394 [cs, q-bio]**, 27 Jul. 2020.

STANTCHEV, G.; DORLAND, W.; GUMEROV, N. Fast parallel Particle-To-Grid interpolation for plasma PIC simulations on the GPU. **Journal of Parallel and Distributed Computing**, v. 68, n. 10, p. 1339–1349, Oct. 2008.

TIAN, X.; BENKRID, K. High-Performance Quasi-Monte Carlo Financial Simulation: FPGA vs. GPP vs. GPU. **ACM Transactions on Reconfigurable Technology and Systems**, v. 3, n. 4, p. 26:1-26:22, 1 Nov. 2010.

VERBONCOEUR, J. P. Particle simulation of plasmas: review and advances. **Plasma Physics and Controlled Fusion**, v. 47, n. 5A, p. A231–A260, 1 May 2005.

VILLASENOR, J.; BUNEMAN, O. Rigorous charge conservation for local electromagnetic field solvers. **Computer Physics Communications**, v. 69, n. 2–3, p. 306–316, Mar. 1992.

VOLKOV, V. **Understanding Latency Hiding on GPUs**. University of California: UC Berkeley, 12 Aug. 2016.

WANG, C. et al. DLAU: A Scalable Deep Learning Accelerator Unit on FPGA. **IEEE Transactions on Computer-Aided Design of Integrated Circuits and Systems**, v. 36, n. 3, p. 513–517, Mar. 2017.

WEIBEL, E. S. Spontaneously Growing Transverse Waves in a Plasma Due to an Anisotropic Velocity Distribution. **Physical Review Letters**, v. 2, n. 3, p. 83–84, 1 Feb. 1959.

XILINX. **Vivado Design Suite User Guide: High-Level Synthesis (UG902)**. URL: <https://www.xilinx.com/support/documentation/sw_manuals/xilinx2017_4/ug902-vivado-high-level-synthesis.pdf>. Accessed in: 31 jan. 2021.

XILINX. **Vivado Design Suite**. URL: <https://www.xilinx.com/products/design-tools/vivado.html>. Accessed in: 30 jan. 2021.

ZOHOURI, H. R. **High Performance Computing with FPGAs and OpenCL**. Tokyo: Tokyo Institute of Technology, Aug. 2018.